\documentclass[twocolumn,aps,pre]{revtex4}
\usepackage{graphicx,color}
\usepackage{pdfpages}
\usepackage{amssymb,amsfonts,amsmath,bm,hyperref,subfig}
\usepackage[capitalise,nameinlink,noabbrev]{cleveref}
\usepackage[latin1]{inputenc}
\usepackage[T1]{fontenc}
\usepackage{makeidx}
\usepackage{graphicx}
\usepackage{chemfig}
\hypersetup{
	colorlinks,
	linkcolor={blue},
	citecolor={blue},
	urlcolor={black}
}
\usepackage{comment}

\begin{document}
    \setlength{\parskip}{0pt}
\title{\bf{Small RNA driven feed-forward loop: Fine-tuning of  protein synthesis through sRNA mediated cross-talk}}

\author{Swathi Tej${^1}$ and Sutapa Mukherji${^2}$${}$}

\email{${^1}$ swathi.rf0897@cftri.res.in\\ ${^2}$ sutapa@cftri.res.in}
\affiliation{$^{}$Protein Chemistry and Technology, Central Food Technological Research Institute, Mysore 570 020, Karnataka, India}

\date{\today}
\begin{abstract}
Often in bacterial regulatory networks, small non-coding RNAs (sRNA) interact with several  mRNA species.
The competition among  mRNAs  for binding to the common pool of sRNA might lead to an effective interaction (cross-talk) between the mRNAs. 
This is similar to the competing endogenous RNA (ceRNA)  effect  wherein the competition to bind to the same pool of micro-RNA 
in Eukaryotes leads to micro-RNA mediated cross-talk resulting in subtle and complex  gene regulation with  stabilised gene expression.  
Here, we study an sRNA-driven feed-forward loop (sFFL) where the top-tier regulator, an sRNA,  binds with two species 
of mRNA for their translational up-regulation. In general, in a  feed-forward loop, the top-tier regulator regulates 
the target protein synthesis through two path ways; while one involves  a direct  regulation of the target protein, 
the other involves  an indirect regulation via up- or down-regulation of an intermediate regulator of the target protein. 
In the present sFFL, an  sRNA (RprA) translationally activates the target protein (RicI) directly and also, indirectly, 
via  up-regulation of its   transcriptional activator ($\sigma^s$). We show that the sRNA-mediated cross-talk between the  
two mRNA species  leads to maximum target protein synthesis for  low synthesis rates of $\sigma^s$-mRNA.
This indicates the possibility of  an optimal target protein synthesis with an efficient  utilisation of $\sigma^s$-mRNA which is typically 
associated with various other stress response activities inside the cell. Since  gene expression is inherently stochastic 
due to the probabilistic nature of various molecular interactions associated with it, we next  quantify the fluctuations in the 
target protein level using   generating function based approach and stochastic simulations. 
The coefficient of variation that provides a measure of fluctuations in the concentration  shows a minimum  
under conditions that also correspond to optimal target protein synthesis. This prompts us to conclude  
that, in sFFL, the  cross-talk leads to optimal target protein synthesis with minimal noise and an efficient utilisation of $\sigma^s$-mRNA.
 \end{abstract}
\maketitle

\section{Introduction}
The gene expression is    a  complex process involving, for example, a number of genes, 
proteins and regulatory  RNAs of different kinds. Different types of molecular mechanism at 
transcriptional and post-transcriptional levels regulate the gene expression process to ensure that 
the proteins are synthesized  to  the desired level with the required efficiency. 
Among various  regulatory molecules, small noncoding RNAs (sRNAs)  have drawn significant attention 
 in the recent past for their  diverse regulatory properties \cite{storz,gottesman2}. 
 sRNAs are  approximately $50-300$ nucleotides long and  
they often regulate the protein synthesis by base-pairing with the target
mRNAs  leading to  mRNA degradation or translational inhibition \cite{gottesman2,papenfort3}.  
Although sRNA mediated 
interactions are found to be mostly repressing in nature, there are recent reports of 
activating interactions by sRNAs 
where sRNAs  enhance the stability of  mRNAs  by sequestering the RNase E recognition site or facilitate translation 
initiation by opening the sequestered ribosome binding site of the secondary structure mRNAs \cite{gottesman1,papenfort3}.
Since sRNAs do not code for proteins, it is generally believed that sRNAs  
 lead to fast and energy-efficient gene regulation. 
 
A number of interesting properties are found in the case of sRNA mediated gene regulation.  For instance, in the case of  only 
sRNA mediated regulation,  the target protein concentration  shows a 
threshold linear response \cite{hwa}.  In the case of combined transcriptional and translational regulation by  proteins 
and  sRNAs, respectively, 
one, however, finds  both monostable (with threshold linear response) and bistable regions  in the target protein concentration \cite{bose}.  
 In addition, sRNAs (or micro-RNAs (miRNA) in case of Eukaryotes) also have  the ability to filter gene expression noise;   
 a property  that might be particularly beneficial for reliable functioning of the cell. 
The gene expression is inherently stochastic  due to  the probabilistic 
 nature of various molecular interactions associated with  gene expression \cite{raj,madanbabu}.
The stochasticity leads to random fluctuations in the protein levels although it is known that   
many  biological functions of the cell require fine-tuning of necessary protein levels. 
 In this regard,  it has been found that sRNA mediated 
repression of gene expression leads to reduced fluctuations in the protein level as compared to transcription factor mediated 
repression \cite{hwa}.  Further,  optimal noise buffering  has  also been seen in more complex 
genetic circuits such as incoherent sRNA mediated feed-forward loops (FFL) \cite{shimoni,osella}, 
sRNA-driven feed-forward loop \cite{swathi} etc.,
where dual strategies i.e. regulation at both transcriptional and translational levels  are employed. 

Recent studies have revealed a particularly   subtle and  complex strategy of   miRNA
mediated gene regulation wherein a given species of  miRNA   interacts with a  number of different 
mRNA targets \cite{bosia1,bossi,marinari,martinoplos,marinari1}. 
 The primary goal of such miRNA-mRNA network is to give rise to regulation through competition in which 
different mRNA targets compete for binding to the same miRNA species (also known as competing 
endogenous RNA or ceRNA effect). In other words, the miRNAs may 
function as a channel  through 
which the change in concentration level in one type of target mRNA  can be conveyed  to another. It 
has been shown that such indirect
miRNA mediated cross-talk may outperform direct regulation under certain circumstances \cite{martinoplos}. 
Additionally,  it has been found  that 
such miRNA mediated cross-talk  between mRNAs results in a broad impact on the protein level
such as enhancing the stability of highly expressed proteins,
altering the correlation patterns of coregulated interacting proteins apart from, in general, 
 fine-tuning the protein levels \cite{marinari1}.

\begin{widetext}
\begin{figure}[hbt]
	\centering
	\fbox{
		\begin{minipage}{7in}
			\includegraphics[width=1\textwidth]{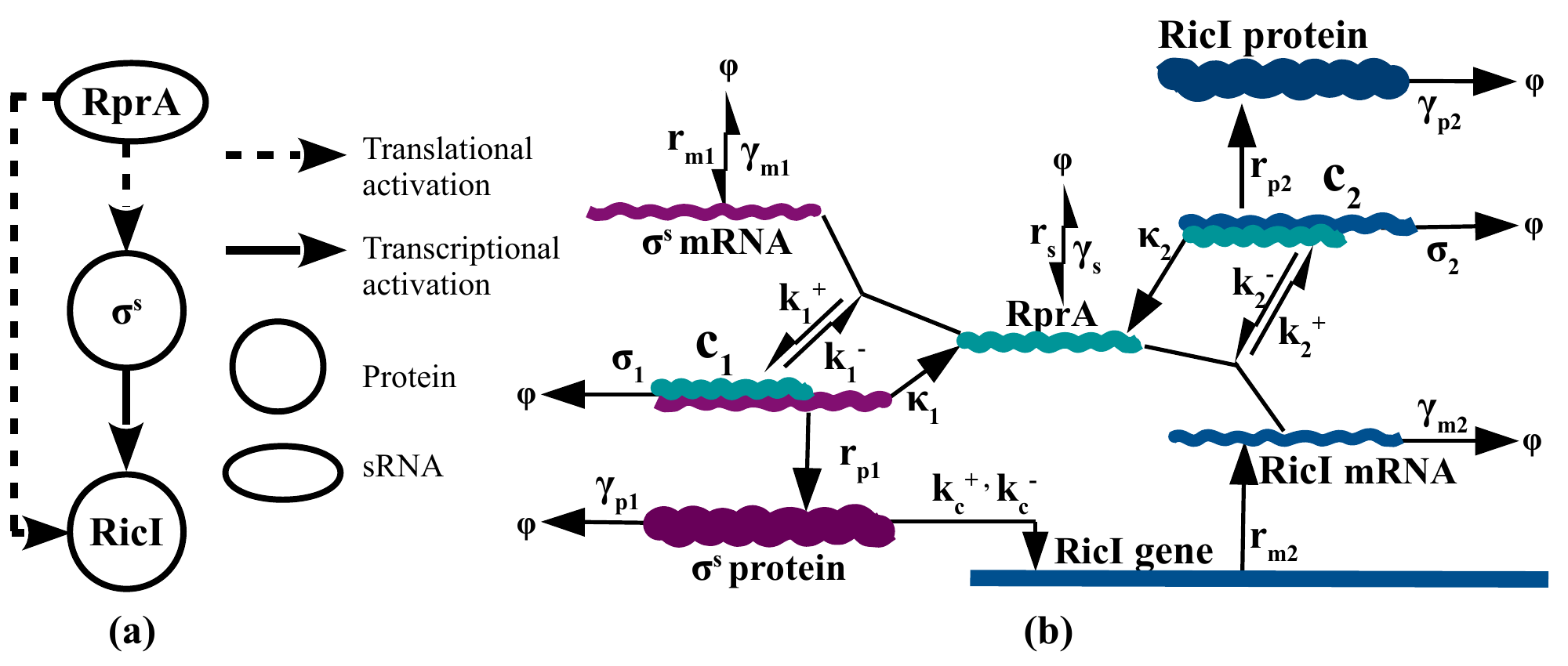}
			\caption{{(a) An sRNA-driven coherent feed-forward loop (sFFL) found in the {\it Salmonella enterica}.
					The top-tier regulator is an sRNA, RprA, which leads to translational up-regulation of  two proteins i.e., $\sigma^s$  and RicI. 
					(b) A  diagram with the details of all the processes such as  synthesis, degradation of all the components as 
					well as the  complex formation between sRNA and mRNAs.  $c_1$ and $c_2$ denote the 
					complexes RprA-$\sigma^s$ mRNA and 
					RprA-RicI mRNA, respectively.}}
			\label{fig:fig-sffl}
		\end{minipage}	}
\end{figure}
\end{widetext}

 In the present paper, our attention is  on a network motif where    the sRNA mediated cross-talk between mRNAs 
 seems to play an important role in  regulating the target   protein synthesis. Network motifs are   
 specific  sub-networks that have frequent recurrences in large regulatory networks 
 as some of their major building blocks \cite{alon}. Such motifs usually have distinct functionalities and it is believed that such motifs 
 are chosen evolutionarily due to distinct advantages  they provide to the cell. 
 The network motif  of our  
 interest  is  a feed-forward loop (FFL) which  is driven by an sRNA (see figure \ref{fig:fig-sffl}(a)) unlike 
 the commonly found FFLs driven by   transcription   factors \cite{papenfort2}.
We refer to this sRNA-driven FFL as sFFL in the following.  In general, 
    the top-tier    regulator in FFL drives the  
 target protein synthesis along two pathways; one pathway involves a direct  
 regulation of the target protein synthesis 
 and the other involves an indirect regulation  via up- or down-regulation of an intermediate 
 regulator of the target protein. For a purely transcriptional FFL, all the interactions 
 are at the transcription level  and the top-tier as well as the 
 intermediate regulators are transcription factors. 
  The sFFL introduced here is different from another kind  of transcription factor driven 
 FFLs which involve sRNA as an intermediate regulator of the target protein \cite{osella,caselle,caselle2}.   
 The  feature that remains common in 
 all these FFLs is that the  top-tier regulator regulates expression of two 
 distinct genes. In the case of sFFL, an sRNA being the top-tier regulator, regulates 
 the translational activities of two different mRNAs along  two regulatory pathways 
 of FFL and thus gives rise to distinct regulatory features through sRNA induced 
 cross-talk between mRNAs.

 The existence  of such  sFFL   has been found experimentally 
very recently in the context of stress response of {\it Salmonella enterica} subjected to stress 
due to a bactericidal agent, bile salt \cite{papenfort2}.     
In this  sFFL, an sRNA, RprA, activates  the synthesis of the target protein, RicI,  directly through 
  translational activation of  RicI mRNA  and indirectly via translational activation in the synthesis of 
  the alternative sigma-factor, $\sigma^s$, which  transcriptionally  
  activates  RicI gene (see figure {\ref{fig:fig-sffl}(a)} and {\ref{fig:fig-sffl}(b)}).  
  By base-pairing with the 
  $\sigma^s$ mRNA,  RprA  opens up the  translation inhibitory 
 loop in the 5$'$ untranslated region  (UTR) of   $\sigma^s$-mRNA   and 
 facilitates ribosome binding  to initiate translation leading to the synthesis of $\sigma^s$ protein \cite{gottesman3}.  
 $\sigma^s$ protein 
 being a  transcriptional activator of RicI gene  leads to an enhanced synthesis   of RicI-mRNA
 which are, then,  translationally activated by sRNA, RprA, resulting in  an  up-regulation in 
   RicI protein synthesis. Here again, RprA facilitates ribosome binding by opening 
   up the translation inhibitory loop of RicI-mRNA.
  Such an FFL with activating regulation 
  along both the paths is known as coherent FFL while an  FFL  involving opposing kind of 
  regulation along the two paths is known as incoherent FFL. An 
  up-regulation of RicI protein happens  in response to the membrane damaging activity of the bile salt. 
  As a response to the stress, the bacterial cell prefers to shut down  the  energy-expensive processes 
  associated with horizontal gene transfer although  horizontal gene transfer plays a significant role 
  in bacteria's survival under normal conditions. 
    By interfering with   the formation of the pilus that is necessary for bacterial 
  conjugation during horizontal gene transfer, RicI protein down-regulates the process of horizontal gene transfer 
    and  protects  the bacteria from additional energy loss associated with  this process
     \cite{papenfort2}. For the rest of the paper, we follow  general notations for various regulatory molecules 
  as listed in table \ref{table-notation}.   
\begin{table}
\caption{Mathematical notations for concentrations of various components of sFFL}
\begin{tabular} {l l l}
	\hline
	Notations 			 				    	&\qquad	\		& Biological sFFL 							\\ \hline \hline
	$s$\ \ \ (sRNA)        			     		&\qquad	\		& RprA sRNA 								\\ 
	$m_1$  (${\rm mRNA1}$)     	    			&\qquad	\	    & $\sigma^s$-mRNA 							\\
	$c_1$\ \ (${\rm complex}\mbox{-}{\rm 1}$)   &\qquad \       &${\rm RprA}\mbox{-}\sigma^s$-mRNA complex  \\  
	$p_1$\ \ (${\rm protein}\mbox{-}{\rm 1}$ )	&\qquad	\ 		& $\sigma^s$ protein  		    			\\ 
	$m_2$\ (${\rm mRNA2}$ )		    			&\qquad	\		& RicI-mRNA 								\\
	$c_2$\ \ (${\rm complex}\mbox{-}{\rm 2}$)   &\qquad \       &${\rm RprA}\mbox{-}{\rm RicI}$-mRNA complex\\ 
	$p_2$\ \ (${\rm protein}\mbox{-}{\rm 2}$) 	&\qquad	\	    &RicI protein 								\\ \hline 
\end{tabular}\label{table-notation}\\
\end{table}

  The miRNA mediated  cross-talk between different mRNAs 
   is a  subject of extensive investigations 
  currently.  Some of the earlier  studies were based on simplified models involving different species of mRNA molecules 
  that were co-repressed by a single miRNA species \cite{marinari}.  The  effect of   miRNA mediated  
   cross-talk on the protein products 
  of these mRNA targets and on the protein-protein interaction were also studied later  starting with this basic framework \cite{marinari1}. 
   In view of the earlier studies,  sFFL considered here is somewhat special for the following reasons.  
  Instead of sRNA mediated repression, which is the most common form of regulation by sRNAs, 
  here the sRNA  leads to translational activation of 
  two different mRNAs.  Further,  the FFL not only  involves sRNA mediated interactions 
  between the mRNAs, but it also involves 
  a direct interaction between the mRNAs. The latter is due to the fact that an 
  up-regulation of one species of mRNA   ($m_1$ i.e. $\sigma^s$-mRNA)  gives rise to  enhanced transcription
   of the other species of mRNA ($m_2$ i.e.  RicI-mRNA)  via up-regulation of   the protein product of $m_1$ 
   (i.e. $\sigma^s$ protein) \cite{papenfort2}. 
  Thus, the sFFL provides a unique platform  to 
  study how the combined  effect of  sRNA mediated  cross-talk and  a direct 
  interaction between the mRNAs influence the target protein regulation.

Using various tools of  mathematical modelling, we quantify the effect of sRNA 
mediated cross-talk on the target protein concentration. 
 The cross-talk  is  seen  by studying how the concentration of mRNA2 ($m_2$) and sRNA ($s$) change as the synthesis rate of 
 mRNA1 ($m_1$) is changed.  Following the table, we refer  the mRNAs as well as their concentrations as $m_1$ and $m_2$. 
 The same convention is followed for the two species of proteins. 
 As the concentration of $m_1$ increases, there is a reduction in the free sRNA concentration 
 since $m_1$ molecules
 tend to form complexes with available free sRNAs and this, consequently, leads to  an increase in  the concentration of 
 free $m_2$ molecules  since 
 sRNAs are largely bound to $m_1$. Interestingly, the change in the 
 concentrations shows a sensitive (or  susceptible) region over a narrow 
 range of $m_1$ synthesis rate ($r_{m_1}$)  where, with a small increase in $r_{m_1}$, there is an abrupt change in 
 various concentrations.  
The  major findings of this study are as follows. 
(1) The sRNA mediated  cross-talk leads to an initial increase in the target protein 
concentration with  the synthesis rate, $r_{m_1}$. With further 
increase in  $r_{m_1}$, the target protein concentration reaches  a peak  and then undergoes  a sharp decrease. 
Hence, it appears that the network motif 
 is designed to perform most efficiently when the supply of $m_1$ (i.e.  $\sigma^s$-mRNA for this sFFL) is low. 
 Since $\sigma^s$ is a key stress 
 response regulator responsible for  various other regulatory activities, this might be an efficient method for maximum 
 utilisation of available $\sigma^s$-mRNA. 
(2) The sRNA mediated cross-talk plays a more important role in regulating the target protein concentration compared to the 
direct interaction between $m_1$ and $m_2$. 
(3) We  show that the range of $r_{m_1}$ over which the maximum synthesis of the target protein takes place also 
corresponds to the range where the noise in the target protein concentration is minimal.   This result is further supported by  
stochastic simulations based on Gillespie algorithm. 
Overall, the present work  suggests  that  the sRNA mediated cross-talk not only  ensures  maximum   target protein synthesis    
with an efficient  use of  $\sigma^s$-mRNA, it also contributes to  maximal noise attenuation in the target protein concentration 
during its synthesis.

\section{Results}
\subsection{Model and steady-state results}
In the following, we present a model for the sFFL describing how  the concentrations of various regulatory components change with time.   
The equations that we use for our calculations are 
\begin{eqnarray}
\dot s&=&r_s-\gamma_s s-k_1^+ s\ m_1-k_2^+ s\ m_2+\nonumber\\ 
&&(k_1^-+\kappa_1)c_1 +(k_2^-+\kappa_2)c_2\label{diffs},\\
\dot m_1&=&r_{m_1}-\gamma_{m_1} m_1-k_1^+ s\ m_1+k_1^- c_1\label{diffm1},\\
\dot c_1&=&k_1^+ s\  m_1-(k_1^-+\sigma_1+\kappa_1) c_1\label{diffc1},\\
\dot p_1&=&r_{p_1} c_1-\gamma_{p_1} p_1\label{diffp1},\\
\dot m_2&=&\frac{r_{m_2} k_c \ p_1}{1+k_c p_1}-\gamma_{m_2} m_2-k_2^+ s \ m_2+k_2^- c_2\label{diffm2},\\
\dot  c_2&=&k_2^+ s\ m_2-(k_2^-+\sigma_2+\kappa_2)c_2 \label{diffc2}, \\
\dot p_2&=&r_{p_2} c_2-\gamma_{p_2} p_2, \label{diffp2}
\end{eqnarray}
where, in general, $\dot x=\frac{d}{dt}x$.  
Here, we have used general notations for various concentrations  as listed in table \ref{table-notation}.
The  rate parameters  $r$ and $\gamma$ represent synthesis and degradation rates in general. $c_1$ and $c_2$  represent 
sRNA-mRNA complexes of two different kinds. $k^+$ and $k^-$ represent the association and dissociation rates of these 
complexes. $\kappa$ and $\sigma$ represent catalytic and stoichiometric
degradation rates of the complexes,  respectively. 
While sRNAs are reused upon catalytic degradation, in the case of stoichiometric degradation the complex is degraded completely. 

\begin{figure}
 \includegraphics[width=0.5\textwidth]{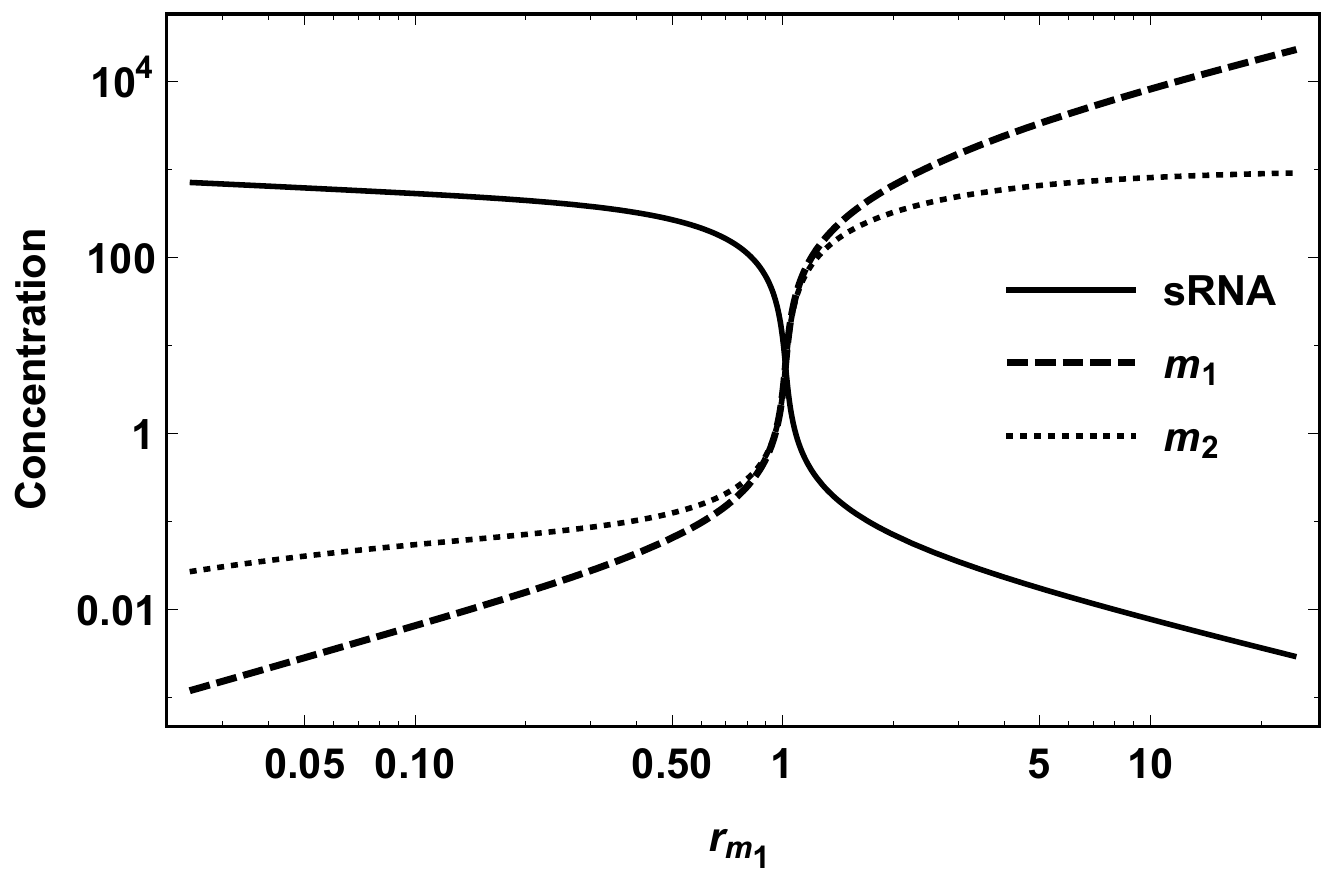}
 \caption{Changes in various concentrations with $r_{m_1}$. The parameter values used 
 for these plots are: $r_s=r_{m_2}=1$, $k_1^+=k_2^+=0.1$, $k_1^-=k_2^-=0.05$, $\sigma_1=\sigma_2=\kappa_1=\kappa_2=0.01$ and $k_c=0.1$. 
 Degradation rates of sRNA, mRNA $m_1$ and $m_2$, proteins $p_1$ and $p_2$   
 are $0.001$ and  the  synthesis  rates  of   $p_1$ and $p_2$ are $0.01$.}
 \label{fig:allconc}
 \end{figure}

  In the steady-state, the  solutions for $m_1$, $m_2$ and $s$ are 
\begin{align}
&m_1=m_1^* F_1(s),\ m_2=\frac{m_2^*\, k_c a\, s\,  F_1(s)\, F_2(s)}{1+k_c\, a\, s\, F_1(s)},\label{mrnas}\ {\rm where}\\
&F_1(s)=\frac{1}{1+s/s_{01}}\ {\rm with}\ s_{01}=\frac{\gamma_{m_1}}{k_1^+}\frac{k_1^-+\sigma_1+\kappa_1}{\sigma_1+\kappa_1},\\
&F_2(s)=\frac{1}{1+s/s_{02}}\ {\rm with}\ s_{02}=\frac{\gamma_{m_2}}{k_2^+}\frac{k_2^-+\sigma_2+\kappa_2}{\sigma_2+\kappa_2},\\
&m_1^*=\frac{r_{m_1}}{\gamma_{m_1}},\   m_2^*=\frac{r_{m_2}}{\gamma_{m_2}},\ a=\frac{r_{p_1}}{\gamma_{p_1}}\frac{k_1^+ m_1^*}{(k_1^-+\sigma_1+\kappa_1)}.
\end{align}
The steady-state concentration of sRNA can be found  upon solving the following algebraic equation for $s$
\begin{align}
r_s-\gamma_s s-r_{m_1}\, \zeta_1\, s\, F_1(s)-
r_{m_2}\, \zeta_2\frac{k_c a\, s^2 F_1(s) F_2(s)}{1+k_c a s F_1(s)}=0, \label{srna1}
\end{align}
\begin{align}
&{\rm with}\  \zeta_1 = \frac{k_1^+ \sigma_1}{\gamma_{m_1}(k_1^- + \sigma_1 +\kappa_1)},\
\zeta_2 = \frac{k_2^+ \sigma_2}{\gamma_{m_2}(k_2^- + \sigma_2 + \kappa_2)}. \  \
\end{align}
Figure \ref{fig:allconc} shows  how sRNA and  mRNA concentrations  change as  $r_{m_1}$,  the transcription rate of $m_1$, is changed.

 The presence of sRNA   induced effective interaction between $m_1$ and $m_2$  is apparent here since a change in the transcription rate of $m_1$ affects the equilibrium concentrations of the sRNA and   $m_2$. 
 Intuitively, as the  concentration of $m_1$ increases, these mRNAs bind the  sRNAs causing a  reduction in  the concentration of free sRNAs and, as 
 a consequence, an increase in the concentration of free $m_2$. Interestingly, 
 figure \ref{fig:allconc}  shows that, over a narrow range of $r_{m_1}$, there exists a sensitive region where 
  the concentrations of s, $m_1$ and $m_2$ change sharply as $r_{m_1}$ is changed. We shall show in the following that 
  over this region, the sFFL attenuates the  fluctuations (noise)  in the target protein concentration maximally. 
In sFFL, there is a direct interaction between mRNAs, $m_1$, and $m_2$,  
 since the protein product of $m_1$ is a transcriptional activator of $p_2$. 
   The change in the concentration of $m_2$ with $r_{m_1}$ is, thus,  expected.   However, as we show below, 
  a   large contribution  in the increase of   $m_2$ concentration  comes from the sRNA induced cross-talk between mRNAs.   
    
\subsection{Response functions}
In order to quantify the interactions among various regulatory molecules, we introduce the following response functions
\begin{eqnarray}
&&\chi_{ij}=\frac{\partial m_i}{\partial  r_{m_j}},\ \ {\rm where }\ \  i,\ j=1,\ 2 \ \ {\rm with}\ \  i\neq j\\
&&\chi_{si}=\frac{\partial s}{\partial r_{m_i}} ,\ \ {\rm where}\ \ i=1, \ 2.
\end{eqnarray}
Here $\chi_{ij}$ represents the  response  in terms of a  change in the concentration of  a specific kind of  mRNA ($m_1$ or $m_2$) 
as  the transcription 
rate  of the other mRNA is changed. In similar way, $\chi_{si}$ represents the change in sRNA concentration as the transcription rates of 
the $i$th  mRNA, i.e. $r_{m_1}$ or $r_{m_2}$ is  changed.

Using (\ref{mrnas}) and (\ref{srna1})  we find the response functions as 
\begin{widetext}
\begin{eqnarray}
\chi_{12} &=& m_1^* F_1' \  \chi_{s2}, \qquad
\chi_{21} = \frac{m_2^*\, k_c a}{{(1 + k_c a s F_1)}^2} \left\lbrace\chi_{s1}\left[ F_1\, F_2\, + s(F_1'F_2 + F_1 F_2')+ k_c\, a\, s^2\, F_1^2\,  F_2'\right]  +\frac{s F_2 F_1}{r_{m_1}} \right\rbrace \ {\rm with } \label{chi-12-21}\\
 \chi_{s1}&=&\frac{-\, \zeta_1\, s\, F_1- \zeta_2\ r_{m_2} k_c\, a\, s^2\, F_2\,  F_1\left[ r_{m_1}(1+ k_c\, a\, s\, F_1)^2\right]^{-1} }{\gamma_s + r_{m_1} \zeta_1 [F_1+ s\, F_1']+ \frac{r_{m_2} \zeta_2 \, k_c a\, s }{(1+ k_c\, a\, s\, F_1)^2}\left[{2\, F_1\, F_2\, + s(F_1'F_2 + F_1 F_2')+ k_c\, a\, s\, F_1^2 (F_2 + s F_2')}\right] } \ {\rm and} \label{dsrm1}\\
\chi_{s2}&=& \frac{- \zeta_2\, k_c a\, s^2\, F_2\,  F_1\left[(1+ k_c\, a\, s\, F_1)\right]^{-1} }{\gamma_s + r_{m_1} \zeta_1 [F_1+ s\, F_1']+ \frac{r_{m_2} \zeta_2 \,  k_c a\, s}{{(1+ k_c\, a\, s\, F_1)}^2} \left[{2\, F_1\, F_2\, + s(F_1'F_2 + F_1 F_2')+ k_c\, a\, s\, F_1^2 (F_2 + s F_2')}\right]}. \label{dsrm2}
\end{eqnarray}	
\end{widetext}
where $F_1'=\frac{\partial}{\partial s} F_1$ and $F_2'=\frac{\partial}{\partial s} F_2$.
The response in the sRNA concentration is always negative ($\frac{\partial s}{\partial r_{m_1}}<0$, $\frac{\partial s}{\partial r_{m_2}}<0$ )  indicating that  the concentration of free sRNA decreases with enhanced synthesis 
of mRNAs, $m_1$ and $m_2$.  Such reduction in free sRNA happens due to the sRNA-mRNA complex formation. 
 The response function $\chi_{21}$ has two parts. The first term    indicates   a change in $m_2$   
as a result of the change in free sRNA concentration. This happens due to an 
 enhanced sRNA-mRNA1 complex ($c_1$) formation   as   the  transcription rate of mRNA $m_1$ is increased.  
  This indirect influence of the transcription rate of $m_1$ on $m_2$ concentration via sRNA  is  a signature of  sRNA induced 
 cross-talk between the mRNAs.   
 The second term in  the equation for  $\chi_{21}$ ( see equation \ref{chi-12-21})
  originates from the direct interaction between $m_1$ and $m_2$ because of  the transcriptional activation of $m_2$ by  
  the protein product of $m_1$. From the equation for  $\chi_{12}$ it is clear that the synthesis rate of  
  $m_2$ affects the concentration of $m_1$ 
  only through an  indirect interaction mediated by sRNA. 
   Further,  as observed earlier \cite{marinari},  in the absence of stoichiometric complex decay (i.e. for $\sigma_1=\sigma_2=0$), 
   the indirect interaction 
term disappears since, in this case, $\frac{\partial s}{\partial r_{m_1}}=0$ and $\frac{\partial s}{\partial r_{m_2}}=0$.  
  The response functions can be compared with sRNA-driven cascade network (sCN) where sRNA 
  post-transcriptionally  up-regulates  $p_1$ synthesis and $p_1$ transcriptionally up-regulates $p_2$ synthesis.
   (see appendix \ref{scn} for details).  
  Thus unlike sFFL,  in sCN,  there is no competition for sRNA sharing.   
  Consequently,  the response functions, in this case,   are  (see appendix \ref{scn} for details)
  \begin{align}
 & \chi_{12}=0,\  \chi_{21}= \frac{k_c m_2^*\,a}{{(1 + k_c a s F_1)}^2} \left\lbrace \chi_{s1}\left[F_1+sF_1'\right]+
  \frac{ s F_1}{r_{m_1}} \right\rbrace\label{chi12-scn}\\
  &{\rm with }\ \   \chi_{s1}=
  \frac{-\, \zeta_1\, s\, F_1}{\gamma_s + r_{m_1} \zeta_1 [F_1+ s\, F_1']}\ \ \  {\rm and} \ \  \ 
   \chi_{s2}=
  0 \label{chis2-scn}
  \end{align}
  The vanishing of $\chi_{12}$ is expected as, in contrary to sFFL,  here the increase in $m_2$ synthesis does not affect the sRNA concentration.
  
 \begin{figure}[b]
		\includegraphics[width=0.4\textwidth]{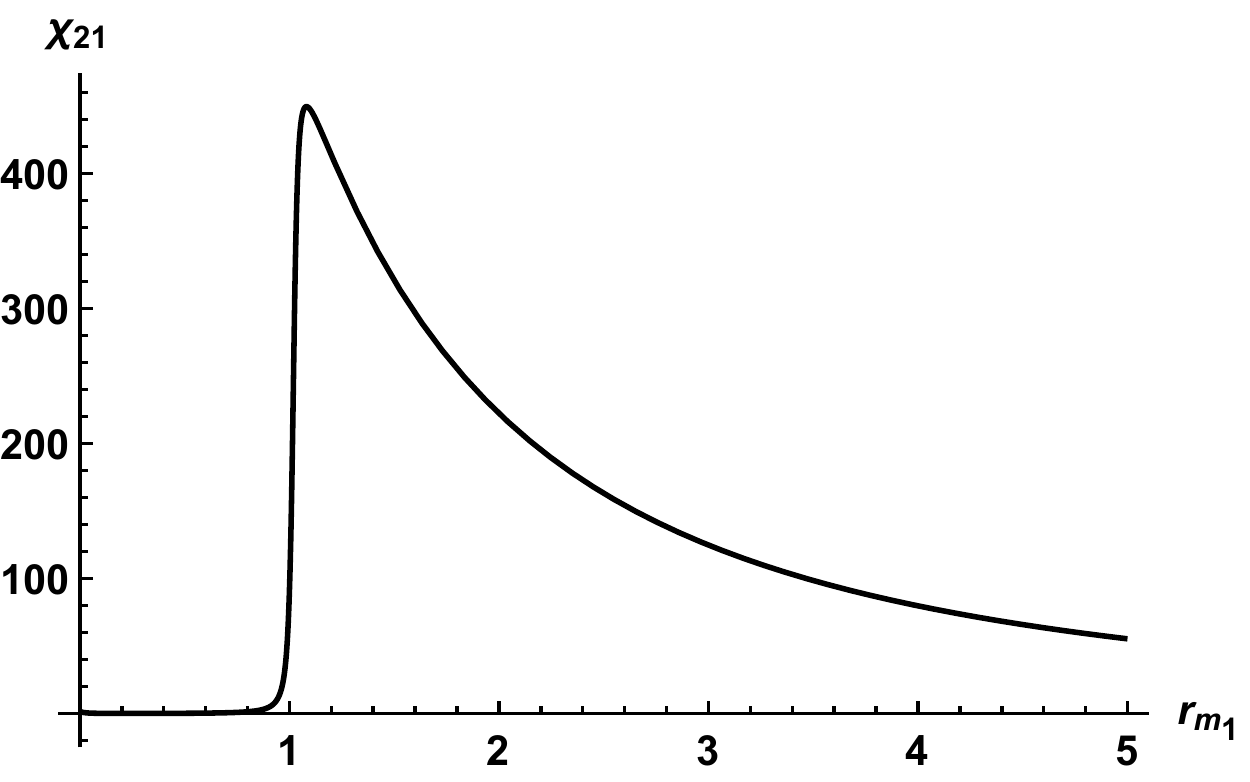} (a)\\
		\includegraphics[width=0.4\textwidth]{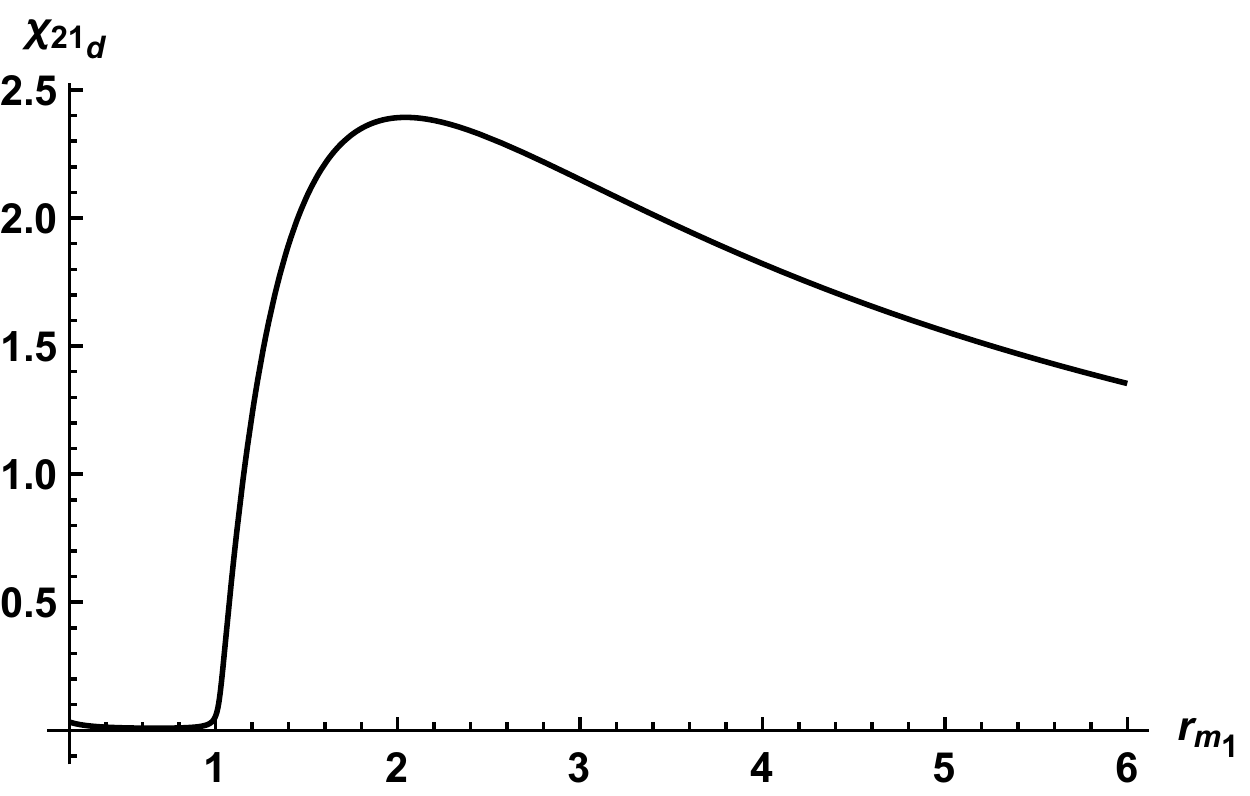} (b)
	\caption{(a) The  response function, $\chi_{21}$ for sFFL  plotted with $r_{m_1}$, the transcription rate of mRNA, $m_1$. (b) The  response due to the transcriptional interaction between $p_1$ (the protein product of $m_1$) and $m_2$.  Same parameter values 
		as in figure \ref{fig:allconc} are used here.  }
	\label{fig:chi}
\end{figure}
 In order to see how  the 
 response function $\chi_{21}$ changes with  $r_{m_1}$, we have obtained  $\chi_{21}$ 
 numerically using equations (\ref{mrnas}) and (\ref{srna1}).
  For an estimate of the contribution from the direct interaction part (the second term in $\chi_{21}$  of equation (\ref{chi-12-21})), $\chi_{21d}$, 
  we have plotted   $\chi_{21d}=m_2^* \frac{ k_c a s F_1 F_2}{r_{m_1}(1+k_c a s F_1)^2}$ after  substituting 
  the solution for $s$ as a function of $r_{m_1}$. 
Figure \ref{fig:chi}(a)  shows  the rate of change of 
   concentration of $m_2$ with respect to  $r_{m_1}$ (i.e. $\chi_{21}$) for different values of $r_{m_1}$. 
  The presence of the sensitive region, as discussed before,  appears very prominently in these figures.  
 As   figures \ref{fig:chi}(a) and \ref{fig:chi}(b) indicate, the  contribution of the direct interaction part to $\chi_{21}$, 
 in general,  is quite small compared to  the sRNA-mediated  part. 


\subsection{Effect of sRNA mediated cross-talk  on the target protein concentration}

 \begin{figure}[h]
	\includegraphics[width=0.4\textwidth]{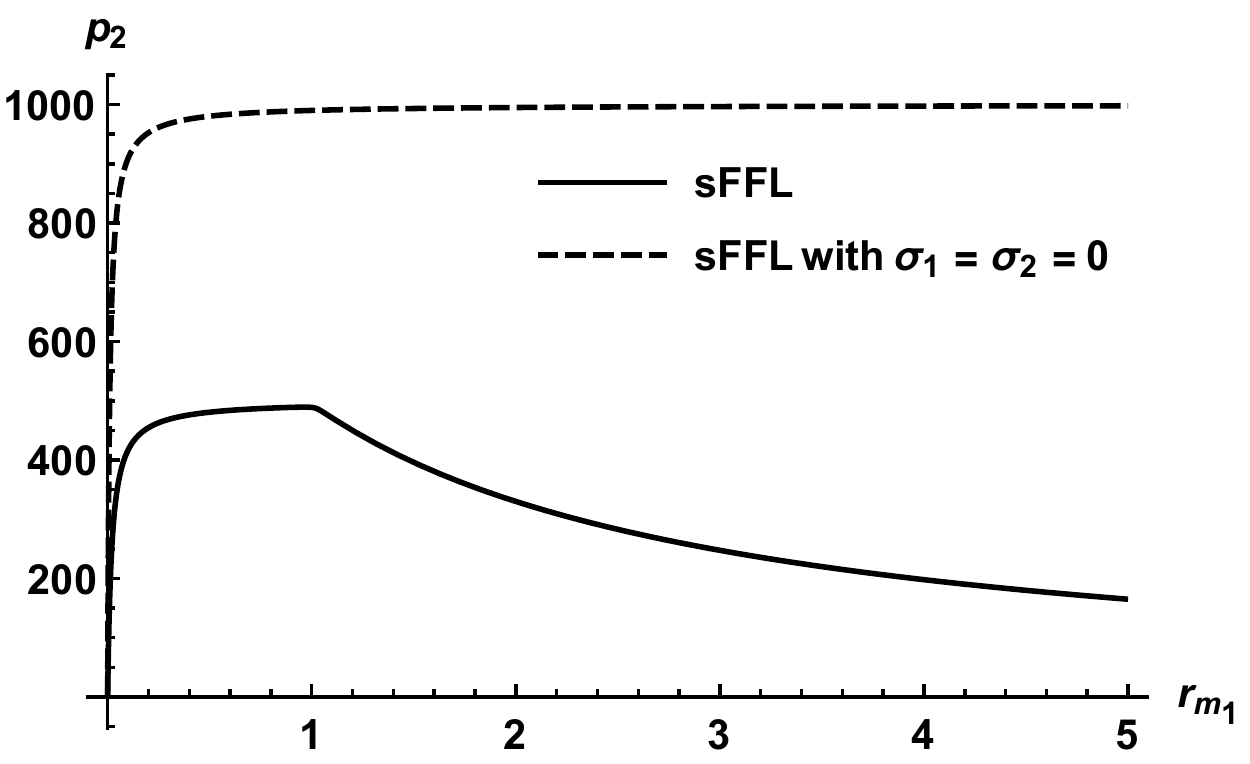}(a)\\
	\includegraphics[width=0.4\textwidth]{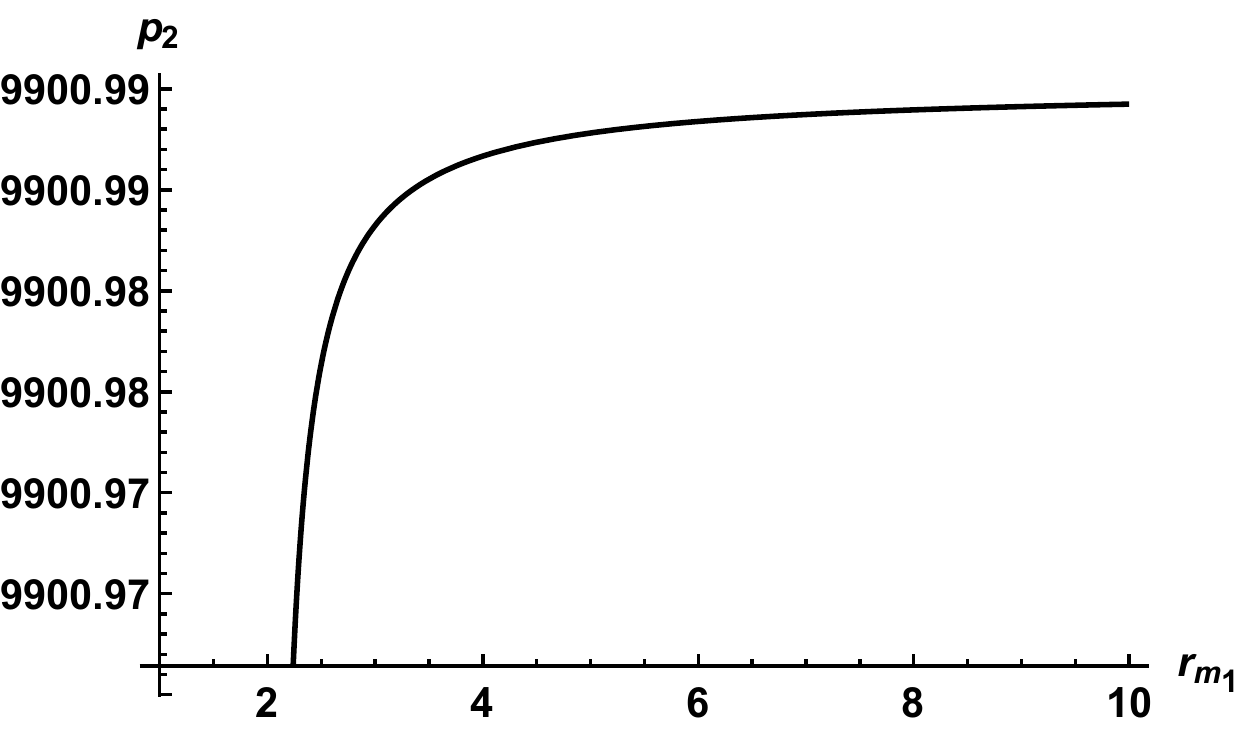}(b)
	\caption{Variation of target protein expression with the transcription rate ($r_{m_1}$) of mRNA, $m_1$ for  (a) sFFL with  and without cross-talk 
		( $\sigma_1 = \sigma_2\, = 0$) and for (b) sRNA-driven  cascade network (sCN).  Parameter values are as same as those chosen for figure \ref{fig:allconc}.}
	\label{fig:p2-sffl-sCN}
\end{figure}

In this section, we focus on the influence of  sRNA mediated  cross-talk  between mRNAs
on the target protein concentration. 
When there is complete  recycling of sRNA (i.e. $\sigma_1=\sigma_2=0$), there is no 
sRNA mediated cross-talk  between the 
mRNAs since  $\chi_{s1}=\chi_{s2}=0$. 
Under such conditions, the 
target protein concentration increases initially with the  transcription rate, $r_{m_1}$,   
and saturates eventually  as a consequence of the 
saturation kinetics associated with  transcriptional activation in  $m_2$ synthesis  by $p_1$ (see figure \ref{fig:p2-sffl-sCN}(a)).  
In the presence of  sRNA induced cross-talk  (i.e.   $\sigma_1, \sigma_2 \neq 0$),  with the increase in $r_{m_1}$, 
the target protein concentration goes through a peak followed by a sharp  decrease. 
 This  happens near the sensitive region where the concentration of free sRNA available for 
 translational activation of $p_2$  drops down drastically. In the case  of 
 sCN  (see figure \ref{fig:p2-sffl-sCN}(b)), the protein concentration reaches a saturation 
 value which persists over the entire range. 
   Thus it appears that  as far as the target protein concentration is concerned,  
   the network performs most efficiently for low concentration  of $\sigma^s$ mRNA. 
   This  feature may be  beneficial for the cell since $\sigma^s$  is a key  
 regulator associated with different types of stress-response of the cell. 

\subsection{Noise Analysis}
\begin{figure}[h!]
	\includegraphics[width=1\linewidth]{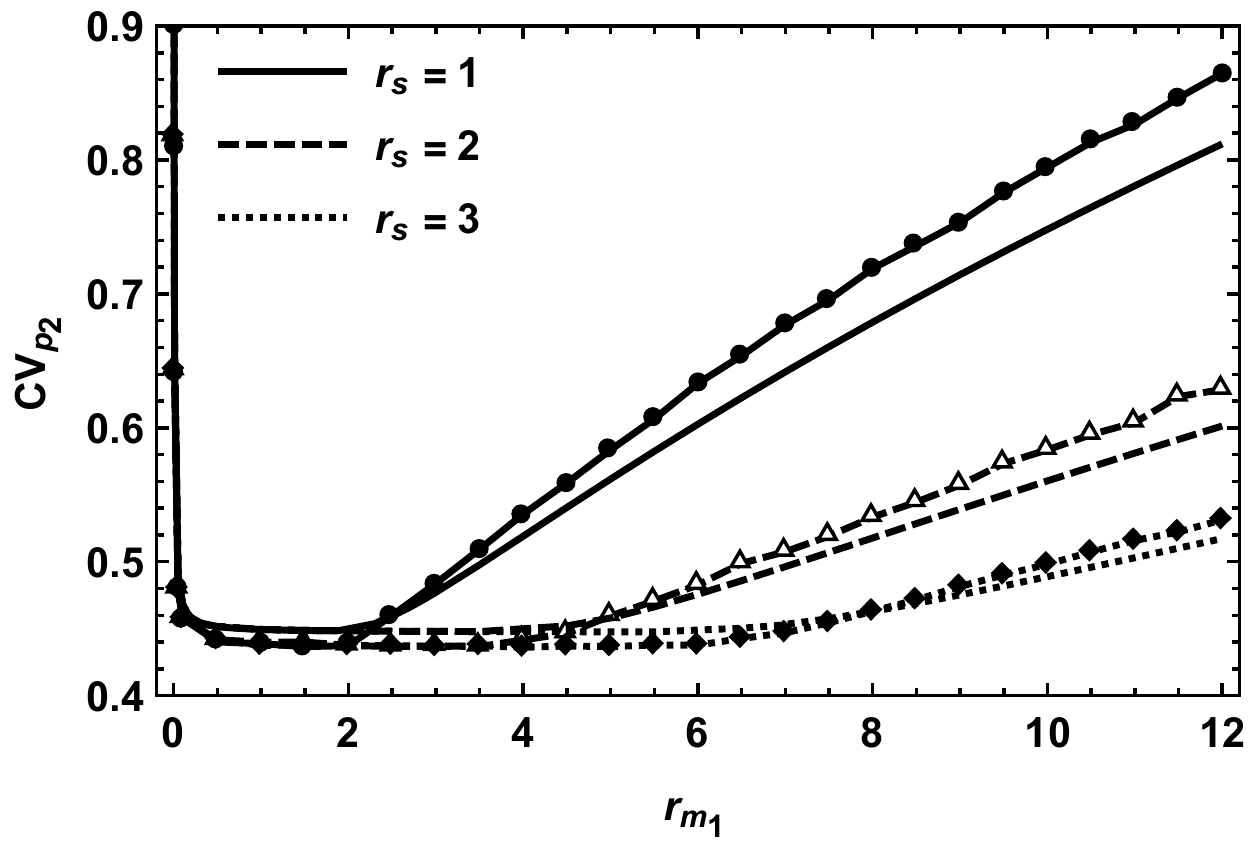}
	\caption{The coefficient of variation for the  target protein  in sFFL  as a function of  $r_{m_1}$, the synthesis rate of the  
		intermediate mRNA, $m_1$, for different values of sRNA synthesis rate, $r_s$. The coefficient of variation is obtained by finding 
		numerical solutions for necessary moments  using Mathematica (lines without markers). The parameter values used 
		for this are 		$\gamma_s=\gamma_{m_1}=\gamma_{m_2}=\gamma_{p_1}=\gamma_{p_2}=0.001$, $r_{m_2}=r_{p_1}=r_{p_2}=0.01,\ 
		k_1^+=k_2^+= 0.1$, $k_1^-=k_2^-=0.05, k_c=0.1$ and  $\kappa_1=\kappa_2=\sigma_1=\sigma_2=0.005$. The lines with markers represent the same found from stochastic simulations based on the Gillespie algorithm. 
		The details regarding the reactions and the reaction rates are presented in appendix \ref{reactions}.  
	} 
	\label{fig:cvp2-rm1}
\end{figure}

The aim of the present section is to find how the sRNA induced cross-talk  influences the noise buffering characteristics of   sFFL. 
We begin with the master equation that describes  how  the probability of a given state changes with time \cite{vankampen}. 
A  state of this system is described by the numbers of sRNA, two different mRNAs and two different protein molecules.  
 The  state of the system  changes  as the numbers of various molecules change due to possible reactions taking place
at given rates. We introduce $P_{s, m_1, p_1, m_2, p_2}(t)$  as the probability at time $t$ that the system  is in a state in which 
the number of sRNA, mRNA and protein molecules are $s$, $m_1$, $m_2$, $p_1$ and $p_2$,  respectively. 
The master equation  describing  the rate of change of the probability function with time can be written as 
\begin{widetext}
	\begin{align}
	&{\partial_t} P_{s,m_1,p_1,m_2,p_2}(t)=r_s (P_{{s-1},m_1,p_1,{m_2},p_2}-P_{s,m_1,p_1,{m_2},p_2}) + \gamma_s ((s+1)P_{s+1,m_1,p_1,{m_2},p_2} - s P_{s,m_1,p_1,{m_2},p_2})\nonumber\\
	&\ \qquad  + r_{m_1} (P_{s,m_1-1, p_1,{m_2},p_2}-P_{s,m_1,p_1,{m_2},p_2}) + \gamma_{m_1} ((m_1+1)P_{s,m_1+1,p_1,{m_2},p_2} - m_1 P_{s,m_1,p_1,{m_2},p_2})\nonumber\\
	&\ \qquad  + r_{p_1}'\ s\ m_1  (P_{s,m_1,p_1-1,{m_2},p_2}-P_{s,m_1,p_1,{m_2},p_2})+ \gamma_{p_1} ((p_1+1)P_{s,m_1,p_1+1,{m_2},p_2} - p_1 P_{s,m_1,p_1,{m_2},p_2})\nonumber\\
	&\ \qquad  + r_{m_2}(p_1) (P_{s,m_1,p_1,{m_2}-1,p_2}-P_{s,m_1,p_1,{m_2},p_2})+ \gamma_{m_2} (({m_2}+1)P_{s,m_1,p_1,{m_2}+1,p_2} - {m_2} P_{s,m_1,p_1,{m_2},p_2})\nonumber \\ 
	&\ \qquad  + r_{p_2}'\ s \ {m_2}\ (P_{s,m_1,p_1,{m_2},{p_2-1}}-P_{s,m_1,p_1,{m_2},p_2}) + \gamma_{p_2} ((p_2+1)P_{s,m_1,p_1,{m_2},p_2+1} - p_2 P_{s,m_1,p_1,{m_2},p_2})\nonumber \\ 
	&\ \qquad  + g_1((s+1)(m_1+1)P_{s+1,m_1+1,p_1,{m_2},p_2}- s\, m_1\, P_{s,m_1,p_1,{m_2},p_2})+  d_1\, s\, ((m_1+1)P_{s,m_1+1,p_1,{m_2},p_2}- m_1\, P_{s,m_1,p_1,{m_2},p_2})\nonumber \\ 
	&\ \qquad  + g_2((s+1)(m_2+1)P_{s+1,m_1,p_1,{m_2+1},p_2}- \ s\, m_2\, P_{s,m_1,p_1,{m_2},p_2})+ d_2\, s\, ((m_2+1)P_{s,m_1,p_1,{m_2+1},p_2}-  m_2\, P_{s,m_1,p_1,{m_2},p_2}).\nonumber \\ \label{mastereqn}
	\end{align} 
	Here ${\partial_t} P_{s,m_1,p_1,m_2,p_2}(t)=\frac{\partial}{\partial t}P_{s,m_1,p_1,m_2,p_2}(t)$.
Various terms on the right hand side of the equation account for different reactions 
representing  synthesis and degradation of various molecules (see appendix \ref{noisesffl} for details). 
Here
\begin{align}
&r_{p_1}'= \frac{r_{p_1}k_1^+}{k_1^- + \sigma_1 + \kappa_1}, \ r_{p_2}'= \frac{r_{p_2}k_2^+}{k_2^- + \sigma_2 + \kappa_2}, \ 
g_1 = \frac{k_1^+\, \sigma_1}{k_1^- + \sigma_1 + \kappa_1},  \  g_2 = \frac{k_2^+\, \sigma_2}{k_2^- + \sigma_2+ \kappa_2}\ \ 
\ d_1 = \frac{k_1^+\, \kappa_1}{k_1^- + \sigma_1 + \kappa_1}, \nonumber \\  & {\rm and}\  d_2 = \frac{k_2^+\, \kappa_2}{k_2^- + \sigma_2 + \kappa_2}.
\end{align}  
The transcriptional activation of $p_2$ by $p_1$ is taken into account by the Hill function, 
$r_{m_2}(p_1)=\frac{r_{m_2} k_c \, p_1}{1+k_c\, p_1}$.
For the present calculation, the Hill function is approximated about the average steady-state density $\langle p_1\rangle$ as 
\begin{eqnarray}
r_{m_2}(p_1)=r_{m_2}^0+r_{m_2}^1 p_1\ {\rm where} \ \ \ r_{m_2}^0 = 
\frac{r_{m_2} k_c^2 \left\langle p_1\right\rangle ^2}{{(1 + k_c  \left\langle p_1\right\rangle) }^2}\   {\rm and} 
\  r_{m_2}^1 = \frac{r_{m_2} k_c}{{(1 + k_c  \left\langle p_1\right\rangle)}^2 }.
\end{eqnarray}
Next, we consider the moment generating function 
\begin{eqnarray}
G(z_1,z_2,z_3,z_4,z_5) =  \sum \limits_{s,m_1,p_1,{m_2},p_2} z^s_1\ z^{m_1}_2\ z^{p_1}_3\ z^{m_2}_4\ z^{p_2}_5\ P_{s,m_1,p_1,{m_2},p_2}\label{eq:mgf}
\end{eqnarray}  whose time evolution is described as 
\begin{eqnarray}
\partial_t G=&&r_s (z_1-1)G+ \gamma_s (1-z_1)\partial_{z_1}G  +r_{m_1}(z_2-1)G+ \gamma_{m_1} (1-z_2)\partial_{z_2}G +r_{p_1}'z_1z_2   (z_3-1)\partial^2_{z_1z_2}G +\gamma_{p_1} (1- z_3)\partial_{z_3}G\nonumber\\
&&+\ r_{m_2}^0 (z_4 -1)G+ r_{m_2}^1 z_3 (z_4  -1) \partial_{z_3}G + \gamma_{m_2} (1 -z_4)\partial_{z_4} G+r_{p_2}'z_1 z_4(z_5 -1)\partial^2_{z_1 z_4}G + \gamma_{p_2} (1 -z_5)\partial_{z_5} G\nonumber\\
&&+ g_1 (1- z_1z_2)\partial^2_{z_1z_2}G + g_2(1- z_1z_4)\partial^2_{z_1z_4}G+d_1z_1(1-z_2)\partial^2_{z_1z_2}G+ d_2 z_1(1-z_4)\partial^2_{z_1z_4}G,  \nonumber\\ \label{diff-generating}
\end{eqnarray}
where $\partial_{z_i}G=\frac{\partial}{\partial z_i} G$ and $\partial^2_{z_i z_j}G=\frac{\partial^2}{\partial z_i \partial z_j} G$.
\end{widetext}

In the  steady state (${\partial_t} G =0$), the  right hand side of  equation (\ref{diff-generating}) is equated  to zero. 
Various average quantities (moments) can be calculated by differentiating the resulting steady-state equation 
with respect to appropriate $z_i$ and then considering $z_i=1$  for all $i$. 
The first moments follow  from $G$ as 
$G_1=\langle s\rangle$, $G_2=\langle m_1\rangle$, $G_3=\langle p_1\rangle$ etc.  where $G_i=\partial _{z_i} G\mid_{\{z_i\}=1}$ with 
$\{z_i\}=1$ indicating   $z_i=1$ for all $i$.
The second moments similarly can be determined from $G_{ij}$ where $G_{ij}=\partial^2 _{z_i z_j}G\mid_{\{z_i\}=1}$. 
We are in particular interested in the target protein fluctuation which can be measured through   the coefficient of variation, 
  $CV_{p_2}=(G_{55}+G_5-G_5^2)^{1/2}/G_5$.   The derivation of the moments $G_5$ and $G_{55}$ appears
 to be complex since the evaluation of a  moment of a given order   involves evaluation of various 
higher order moments.  In order to simplify the derivation, we restrict ourselves up to second moments and express 
the third moments necessary for this derivation  in terms of   lower order moments using Gaussian approximation \cite{lafuerza}. 
The results for moments up to second order and the approximate forms 
of the required third order moments are presented in appendix \ref{moments}. 

Moments required for the evaluation of the coefficient of variation are found numerically using Mathematica. 
The coefficient of variation as a function of $r_{m_1}$ for different values of, $r_s$,  the sRNA synthesis rate, 
 is  shown in figure \ref{fig:cvp2-rm1}. It is clear that the fluctuations in the target 
protein concentration are minimum over a range of $r_{m_1}$. Interestingly, this minimum 
region overlaps with the sensitive region  shown in   figure \ref{fig:allconc}.  As  $r_s$ increases, 
the sensitive region extends towards larger $r_{m_1}$. Figure \ref{fig:cvp2-rm1} shows that   the minimum 
fluctuation region also changes accordingly  and there is a systematic increase in the fluctuation
 beyond this region.  In order to see the role of cross-talk in noise processing 
characteristics, we have also obtained the coefficient of variation for the target protein number for 
sFFL with $\sigma_1=\sigma_2=0$ (no cross-talk) and sCN (see appendix \ref{noisesffl} and \ref{noisescn}). 
The coefficients of variation plotted with $r_{m_1}$  
indicate the absence of optimal noise attenuation.   

In order to verify the above observations, we obtain exact results for the coefficient of variation of the target protein number from stochastic simulations based on Gillespie algorithm \cite{gillespie,gillespie1}. In simulations, we begin with an initial number of various molecular species. The key reactions in our simulation include synthesis, degradation, complex association/dissociation and transcriptional interactions between the DNA and the transcription factor. In stochastic simulations, an event or a  reaction and the interval between two successive reactions are chosen probabilistically. Based on the reaction that takes place, at each simulation step, the number of molecular species is updated accordingly. We allow the system to evolve for $5\times 10^7$ steps and record the target protein number after every $500$  steps leaving about initial $2\times 10^5$ steps. The coefficient of variation averaged over $100$ samples  has been
presented in figure \ref{fig:cvp2-rm1}. A remarkable agreement between results from simulations and from generating function method further confirms that maximum noise attenuation in the target protein level happens over the sensitive region.

\begin{figure}[h!]
	\includegraphics[width=1\linewidth]{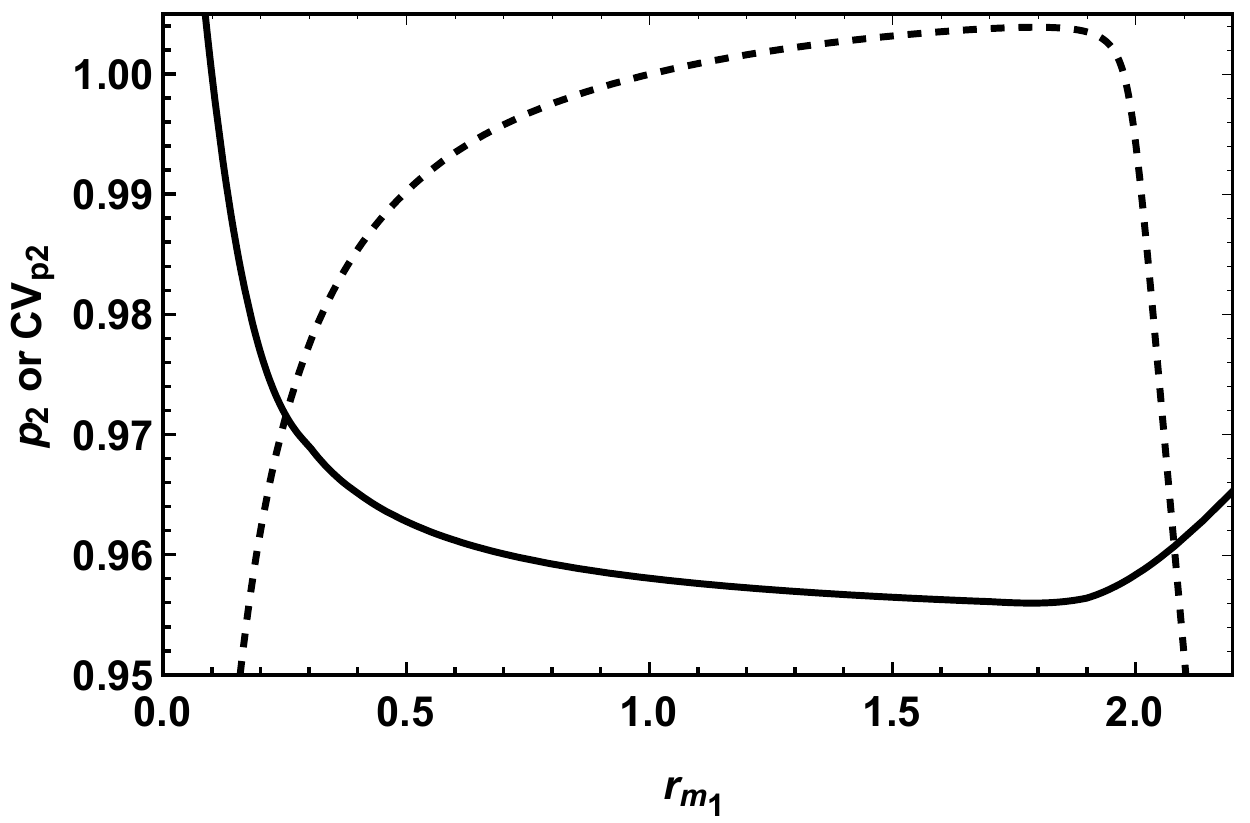}
	\caption{Target protein concentration (dashed line) (normalized with respect to its peak value)  and the coefficient of variation (solid line) plotted with 
		$r_{m_1}$. The coefficient of variation is obtained  from the analysis based on generating functions. The parameter values are same 
		as those used for figure \ref{fig:cvp2-rm1}. For this figure $r_s=1$. } 
	\label{fig:cvp2-overlap}
\end{figure}

Our conclusions on how the target protein concentration as well as its fluctuations around the mean value change with $r_{m_1}$ can be summarized through a single plot shown in   figure \ref{fig:cvp2-overlap}. This plot clearly shows that the range of $r_{m_1}$ over which the target protein concentration reaches its maximum also corresponds to the range of maximum noise attenuation in the target protein level. Thus it appears that  sRNA-mediated cross-talk plays a subtle role in ensuring that maximal target protein synthesis happens with the least noise in the target protein level.

\section{Discussion}
Small noncoding RNAs in bacterial cells  are major regulators driving a number of biological processes  such as stress response,  
biofilm formation, quorum sensing etc. under different kinds of environmental signals.  It is commonly found that  one species of sRNA  can regulate translation of  several  species 
of mRNAs and also a species of mRNA can be targeted by more than one species of sRNAs. The complex network formed thereby 
is believed to be able to integrate  different types of environmental signals in a unified manner.  The regulation here primarily happens 
through  sharing of sRNAs by  several co-regulated  mRNA targets.  If the   concentration of one species 
of mRNA, say $m_1$ increases, the sRNAs are expected to  bind to  $m_1$ predominantly due to its increased concentration. This 
 leads to an increase in sRNA-mRNA1 complex  ($c_1$) concentration and consequently a lowering of  available free 
 sRNAs and an increase 
 in the concentration of other mRNA targets.    The sharing of sRNA thus provides a link between its target mRNAs 
 leading  to an sRNA mediated effective interaction (cross-talk) between the mRNAs. Mostly this link extends over mRNAs from 
 different networks  and the  regulation through sRNA sharing may give  rise to the possibility of integrating biological 
 processes governed by different networks. In the present work, we are interested in understanding the effect of sRNA mediated 
 cross-talk  between two species of mRNAs in a network motif comprised of  feed-forward loop driven by an sRNA. Thus unlike 
 earlier work, our focus is on the effect of sRNA sharing on the target protein synthesis in  a network motif. 
 
 In the  feed-forward loop of our interest, the top-tier regulator is an sRNA (RprA) that regulates the target protein (RicI) 
 synthesis along two  pathways. While one pathway involves a direct translational activation of RicI protein by  RprA, 
  the other pathway  involves an indirect activation of RicI expression  via translational activation of its  transcriptional activator 
  ($\sigma^s$) by  RprA. The sRNA thus translationally activates its two target mRNAs, RicI mRNA and 
  $\sigma^s$ mRNA.  
  Such an sRNA-driven feed-forward loop(sFFL) is found to play a significant role in regulating horizontal gene transfer  in 
 {\it Salmonella enterica} when the bacteria is subjected to stress due to  membrane damaging activities of a bactericidal agent, 
 bile salt. Since horizontal gene transfer is an energy-expensive process,  inhibiting such processes during stress conditions 
 might be a preferred strategy for additional  protection of the bacteria.  RicI protein participates in the stress-response 
 activity by  interfering  with  pilus formation which is necessary for bacterial conjugation during  the process of horizontal gene 
 transfer. 
 
 The $\sigma^s$ protein which is a transcriptional activator of RicI  is known to be a key stress-response regulator  
that  responds to  various other environmental 
 stress conditions too. The up-regulation of $\sigma^s$ protein, however, does not necessarily imply an 
 up-regulation of RicI protein. The  
 design of sFFL ensures   that RicI is up-regulated by combined action along both the paths of sFFL.   
  The present analysis further shows that as a result of 
 sRNA-mediated cross-talk, the network also performs most efficiently when the $\sigma^s$ mRNA concentration is low.  
This result is concluded from the following observations.
If the synthesis rate of $\sigma^s$ mRNA is increased, the concentration of free sRNA  decreases as a consequence 
 of sRNA-mRNA1 complex formation. Due to reduced availability of free sRNA molecules for binding, 
 the level of free $m_2$ increases. 
 This  effect of cross-talk linking the two species of mRNAs  is more abrupt over a narrow  range  of $m_1$ synthesis 
 rate which  we 
 refer as a sensitive region. In this region, 
   a small increase in the synthesis rate  gives rise to  a sharp drop in the sRNA concentration along with 
a steep rise 
   in the concentrations of  the two mRNAs. As far as the target protein  is concerned, 
  it is found that with the increase in $m_1$ synthesis rate, the target protein concentration reaches a maximum 
 and then decreases  sharply  
  as a consequence of a sharp drop in  free sRNA in the sensitive region. Although  a large number of 
  $m_2$ are synthesized, the translation is less probable due to less availability of free sRNA molecules  near  the centre 
  of the  sensitive region.  Such  variation in the target protein concentration indicates    that the motif performs 
 most efficiently for low $\sigma^s$ synthesis rate. Since $\sigma^s$ is a major stress-response regulator 
 associated with different kinds of environmental stress,  such optimal utilization of $\sigma^s$ mRNA might 
 be a beneficial  regulation strategy for  the cell. 
 
  The gene expression is intrinsically  noisy with  noise originating from the  randomness associated with various 
  molecular interactions. This raises a fundamental question as to how  cells perform in a robust manner despite 
 significant variations in the gene expression levels. Recent  evidences suggest that the architecture of the 
 regulatory network  determines the effect of gene expression noise on the target protein level, with some architecture 
 leading to noise filtering,  
 some leading to noise amplification  etc.  It is believed that the network motifs are evolutionarily selected based on the
 network architecture that may lead to such distinct noise processing characteristics.   
 In view of this knowledge, we  attempt to find  the interplay of   cross-talk and gene-expression noise in sFFL and its 
 implications on   fluctuations in the target protein level. 
To this end, we use a generating function based method to obtain the coefficient of variation in the target protein level. 
Plotting the coefficient of variation  with  $\sigma^s$ mRNA synthesis rate, we find 
 that the noise in the target protein level is minimized over the sensitive region. In particular, the range over which 
 maximum noise buffering  happens 
 also coincides with the range over which  maximum target protein synthesis can be achieved. These results are verified 
 through stochastic simulations based on Gillespie algorithm. Thus, it appears  that 
 the network architecture     not only leads to   maximum target protein synthesis under limited $\sigma^s$ mRNA 
 concentration, it also leads to maximal noise buffering while synthesizing the target protein to its maximum level. 
 All these features resulting  primarily from the sRNA mediated cross-talk between mRNAs indicate a complex and 
 precise level of gene regulation through sRNA.

\bigskip
{\bf Conflict of interest:} Authors declare no conflict of interest. 

\bigskip
{\bf Acknowledgment:}  ST and SM   thank DBT, India for financial support through grant no. BT/PR16861/BRB/10/1475/2016. ST thanks University Grants Commission (UGC), India for financial support under CSIR-UCG NET-JRF.

\appendix
\section*{APPENDIX}

\section{sRNA-driven Cascade Network (sCN)}\label{scn}
\begin{figure}[h]
	\centering
	\includegraphics[width=\linewidth]{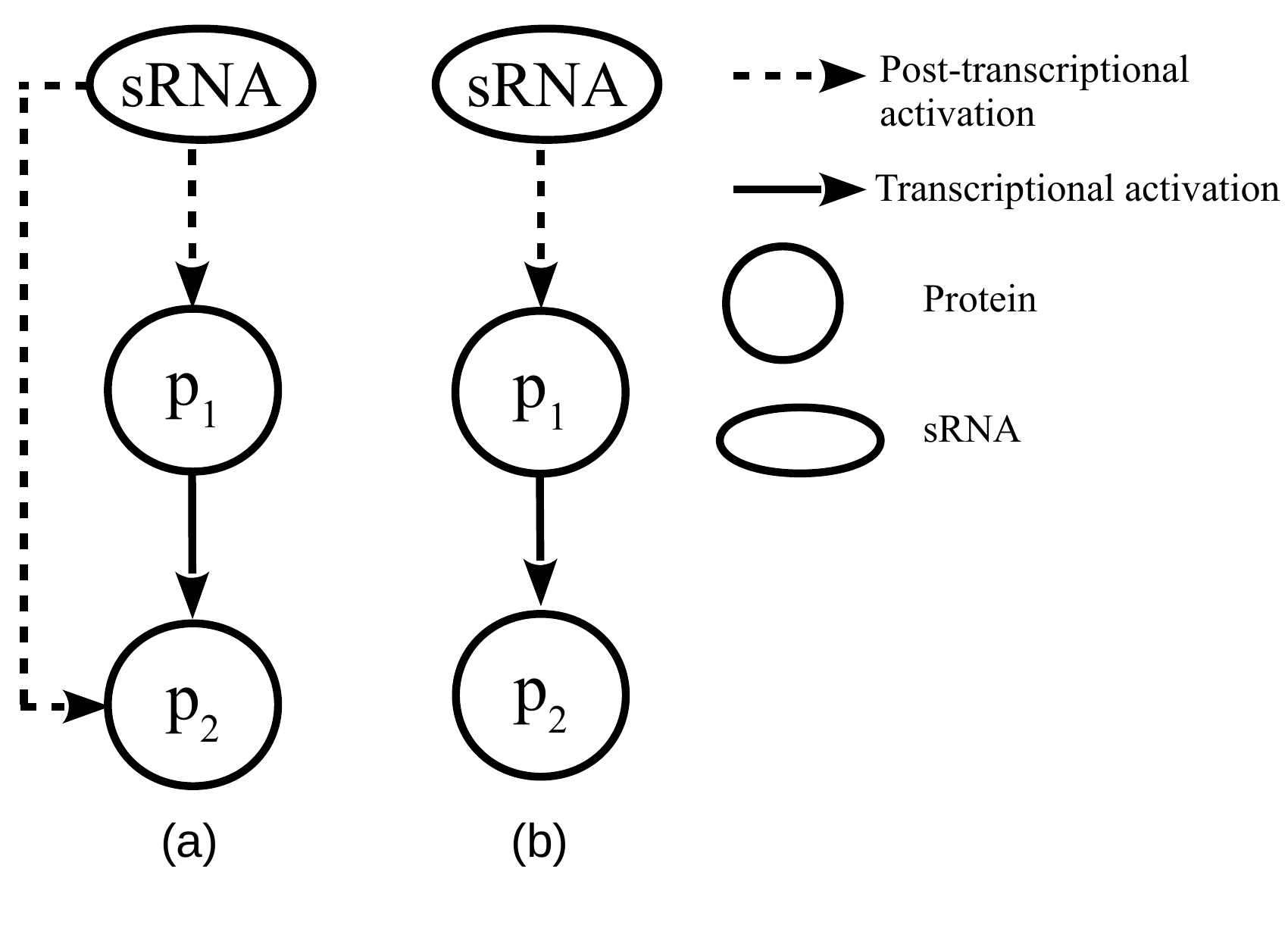}
	\caption{Schematic diagrams for (a) sFFL and (b) sCN}
	\label{fig:scn}
\end{figure}

In this section, we consider the  sRNA-driven cascade network (sCN) where sRNA translationally activates protein $p_1$ which 
then transcriptionally activates protein $p_2$ (see figure \ref{fig:scn}(b)). Thus, it is a cascade network where no sharing of free sRNA takes 
place. 
 The variation in concentrations of  various components of sCN with time is as follows,
\begin{align}
\dot s&=r_s-\gamma_s\, s- k_1^+\, s\, m_1   + (k_1^- + \kappa_1 ) c_1, \\
\dot m_1&=r_{m_1} -\gamma_{m_1}\, m_1 -k_1^+\, s\, m_1 +k_1^-\, c_1,\\
\dot c_1&=k_1^+\, m_1\, s - (k_1^- + \sigma_1 + \kappa_1 ) c_1,\\
\dot p_1&=r_{p_1}\, c_1 -\gamma_{p_1}\, p_1,\\
\dot m_2&=\frac{r_{m_2}\,{k_c}\, p_1}{1+ k_c\, p_1} -\gamma_{m_2}\, m_2,\ \ {\rm and}\\
\dot p_2&=r_{p_2}\, m_2 -\gamma_{p_2}\, p_2.
\end{align}

The steady-state concentrations of mRNAs $m_1$ and $m_2$ in terms of functions $F_1(s)$ and $F_2(s)$ are
\begin{eqnarray}
m_1 &=& m_1^*\, F_1(s),\label{m1scn}\\
m_2 &=& \frac{m_2^*\, k_c\, a\, s\, F_1(s)}{1+ k_c\, a\, s\, F_1(s)} \ \ {\rm where} \label{m2scn}
\end{eqnarray}
$m_1^* =\frac{r_{m_1}}{\gamma_{m_1}},\quad   m_2^*= \frac{r_{m_2}}{\gamma_{m_2}},\ \  \quad  a= \frac{r_{p_1}\, k_1^+\, m_1^*}{\gamma_{p_1} (k_1^-+ \sigma_1  + \kappa_1 )},$ $F_1(s)=\frac{1}{1+s/s_{01}}$ and 
$s_{01}=\frac{\gamma_{m_1}}{k_1^+}\frac{k_1^-+\sigma_1 +\kappa_1}{\sigma_1 +\kappa_1}$. 
In the steady-state, the   concentration of sRNA,  can be found  by solving   $r_s - \gamma_s\, s - r_{m_1}\, \zeta_1\, s\, F_1(s) =0$, 
where $\zeta_1=\frac{k_1^+ \sigma_1}{\gamma_{m_1}(k_1^-+ \sigma_1+\kappa_1 )}$. 
 The target protein concentration  for sCN  is $p_2=\frac{r_{p_2}}{\gamma_{p_2}} m_2$ where $m_2$ is as shown in equation (\ref{m2scn}).



\section{Noise analysis for  sFFL}\label{noisesffl}

\subsection{Equations and notations}
The reactions incorporated into the master equation, equation (\ref{mastereqn}), are based on the following effective equations. 
\begin{eqnarray}
&&\dot{s}=r_s-\gamma_s\, s -g_1\, s\, m_1-g_2\,  s\, m_2,\\
&&\dot{m_1}=r_{m_1}-\gamma_{m_1}\, m_1-d_1 \, s\, m_1-g_1\, s\, m_1,\\
&&\dot{p_1}=r_{p_1}' s\, m_1-\gamma_{p_1}\,  p_1,\\
&&\dot{m_2}=\frac{r_{m_2}\, k_c\, p_1}{1+k_c\, p_1}-\gamma_{m_2}  m_2-d_2\, s\, m_2-g_2\, s\, m_2,  \\
&&{\rm and} \ \ \ \dot{p_2}=r_{p_2}' s\, m_2 -\gamma_{p_2}\,  p_2.
\end{eqnarray}
These equations are 
derived from (\ref{diffm1})-(\ref{diffp2})  by substituting $c_1=\frac{k_1^+\, s\, m_1}{k_1^-+\sigma_1 +\kappa_1} $ and $c_2=\frac{k_2^+\,s\, m_1}{k_2^-+\sigma_2 + \kappa_2}$.
While $g_1 = \frac{k_1^+\, \sigma_1}{k_1^- + \sigma_1 +\kappa_1}  \qquad {\rm and} \qquad g_2 = \frac{k_2^+\, \sigma_2}{k_2^- + \sigma_2 + \kappa_2}$
 correspond to  combined degradation of  sRNA and mRNA, $d_1 = \frac{k_1^+\, \kappa_1}{k_1^- + \sigma_1 + \kappa_1}  \qquad {\rm and} 
 \qquad d_2 = \frac{k_2^+\, \kappa_2}{k_2^- + \sigma_2 + \kappa_2}$ indicate degradation of mRNA alone while sRNAs are recycled back.
 Further, here $r_{p_1}'= \frac{r_{p_1}k_1^+}{k_1^- + \sigma_1 + \kappa_1} \qquad {\rm and} \qquad r_{p_2}'= \frac{r_{p_2}k_2^+}{k_2^- + \sigma_2 + \kappa_2}$.

\subsection{Moments}\label{moments}
In this section, we list the results for  first and second order moments obtained from the generating function approach.  
The results show that moments of a given order involve higher order moments. In order to simplify our calculations, 
we consider moments up to second order and express the  third order moments in terms of lower order moments 
using Gaussian approximation. 
\begin{align}
&G_1 = \frac{r_s - g_1\,G_{12} - g_2\,G_{14}}{\gamma_s}\\
&G_2 = \frac{r_{m_1} - (g_1+ d_1) G_{12}}{\gamma_{m_1}}\\
&G_3 = \frac{r_{p_1}' G_{12}}{\gamma_{p_1}}\\ 
&G_4 = \frac{r_{m_2}^0 + r_{m_2}^1\, G_3 - (g_2+ d_2) G_{14} }{\gamma_{m_2}}\\
&G_5 = \frac{r_{p_2}' G_{14}}{\gamma_{p_2}}\\
&G_{11} = \frac{r_s\, G_1 - g_1\,G_{112} - g_2\,G_{114}}{\gamma_s}\\
&G_{22} = \frac{r_{m_1} G_2 - (g_1+ d_1) G_{122} }{\gamma_{m_1}}\\
&G_{33} = \frac{r_{p_1}' G_{123}}{\gamma_{p_1}}\\ 
&G_{44} = \frac{r_{m_2}^0 G_4 + r_{m_2}^1 G_{34} - (g_2+ d_2) G_{144}  }{\gamma_{m_2}}\\ 
&G_{55} = \frac{r_{p_2}' G_{145}}{\gamma_{p_2}}\\
&\scalebox{1.25}
{$G_{12} = \frac{r_s G_2 + r_{m_1} G_1 - g_1(G_{112} + G_{122}) - g_2G_{124} - d_1G_{112}}{\gamma_s + \gamma_{m_1} + g_1+ d_1}$}\\
&\scalebox{1.25}
{$G_{13} = \frac{r_s G_3 + r_{p_1}' ( G_{12} + G_{112}) - g_1\,G_{123} - g_2G_{134}}{\gamma_s + \gamma_{p_1}}$}\\
&\scalebox{1.1}
{$G_{14} = \frac{r_s G_4 + r_{m_2}^0 G_1 + r_{m_2}^1  G_{13} - g_1\,G_{124} -( g_2 + d_2)G_{114} -g_2 G_{144}}{\gamma_s + \gamma_{m_2} + g_2+ d_2 }$}
\end{align}
\begin{align}
&\scalebox{1.25}
{$G_{15} = \frac{r_s G_5 + r_{p_2}' ( G_{14} + G_{114}) - g_1\,G_{125} - g_2\,G_{145}}{\gamma_s + \gamma_{p_2}}$}\\
&\scalebox{1.25}
{$G_{23} = \frac{r_{m_1} G_3 + r_{p_1}' ( G_{12} + G_{122}) - (g_1+ d_1) G_{123}}{\gamma_{m_1} + \gamma_{p_1}}$}\\
&\scalebox{1.25}
{$G_{24} = \frac{r_{m_1} G_4 + r_{m_2}^0 G_2 + r_{m_2}^1 G_{23} - (g_1+ g_2+ d_1+ d_2 ) G_{124} }{\gamma_{m_1} + \gamma_{m_2}}$}\\
&\scalebox{1.25}
{$G_{25} = \frac{r_{m_1} G_5 + r_{p_2}' G_{124} - (g_1+ d_1) G_{125}}{\gamma_{m_1} + \gamma_{p_2} }$}
\end{align}
\begin{align}
&\scalebox{1.25}
{$G_{34} = \frac{r_{p_1}' G_{124} + r_{m_2}^0  G_3 + r_{m_2}^1 ( G_3 + G_{33}) - (g_2+ d_2) G_{134}}{\gamma_{p_1} + \gamma_{m_2}}$}\\
&\scalebox{1.25}
{$G_{35} = \frac{r_{p_1}' G_{125} + r_{p_2}' G_{134}}{\gamma_{p_1} + \gamma_{p_2} }$}
\end{align}
Using Gaussian approximation, we express  the third order moments  in terms of the lower order moments as shown below. 
\begin{align}
&G_{112} = G_{11}G_{2} + G_{1} G_{2} + 2 G_{1}G_{12} - 2 G_{1}^2G_{2}-G_{12}\\
&G_{113} = G_{11}G_{3} + G_{1} G_{3} + 2 G_{1}G_{13} - 2 G_{1}^2G_{3}-G_{13}\\
&G_{114} = G_{11}G_{4} + G_{1} G_{4} + 2 G_{1}G_{14} - 2 G_{1}^2G_{4}-G_{14}\\
&G_{122} = G_{22}G_{1} + G_{1} G_{2} + 2 G_{2}G_{12} - 2 G_{2}^2G_{1}-G_{12} \\ 
&G_{123} = G_{12}G_{3} + G_{13} G_{2} +  G_{23}G_{1} - 2 G_{1}G_{2}G_{3} \\ 
&G_{124} = G_{12}G_{4} + G_{14} G_{2} +  G_{24}G_{1} - 2 G_{1}G_{2}G_{4}\\ 
&G_{125} = G_{12}G_{5} + G_{15} G_{2} +  G_{25}G_{1} - 2 G_{1}G_{2}G_{5}\\
&G_{133} = G_{33}G_{1} + G_{1} G_{3} + 2 G_{3}G_{13} - 2 G_{3}^2G_{1}-G_{13}\\ 
&G_{134} = G_{13}G_{4} + G_{14} G_{3} +  G_{34}G_{1} - 2 G_{1}G_{3}G_{4}\\  
&G_{135} = G_{13}G_{5} + G_{15} G_{3} +  G_{35}G_{1} - 2 G_{1}G_{3}G_{5}\\
&G_{144} = G_{44}G_{1} + G_{1} G_{4} + 2 G_{4}G_{14} - 2 G_{4}^2G_{1}-G_{14}\\ 
&G_{145} = G_{14}G_{5} + G_{15} G_{4} +  G_{45}G_{1} - 2 G_{1}G_{4}G_{5}
\end{align}

\begin{figure}[h]
	\centering
	\includegraphics[width=\linewidth]{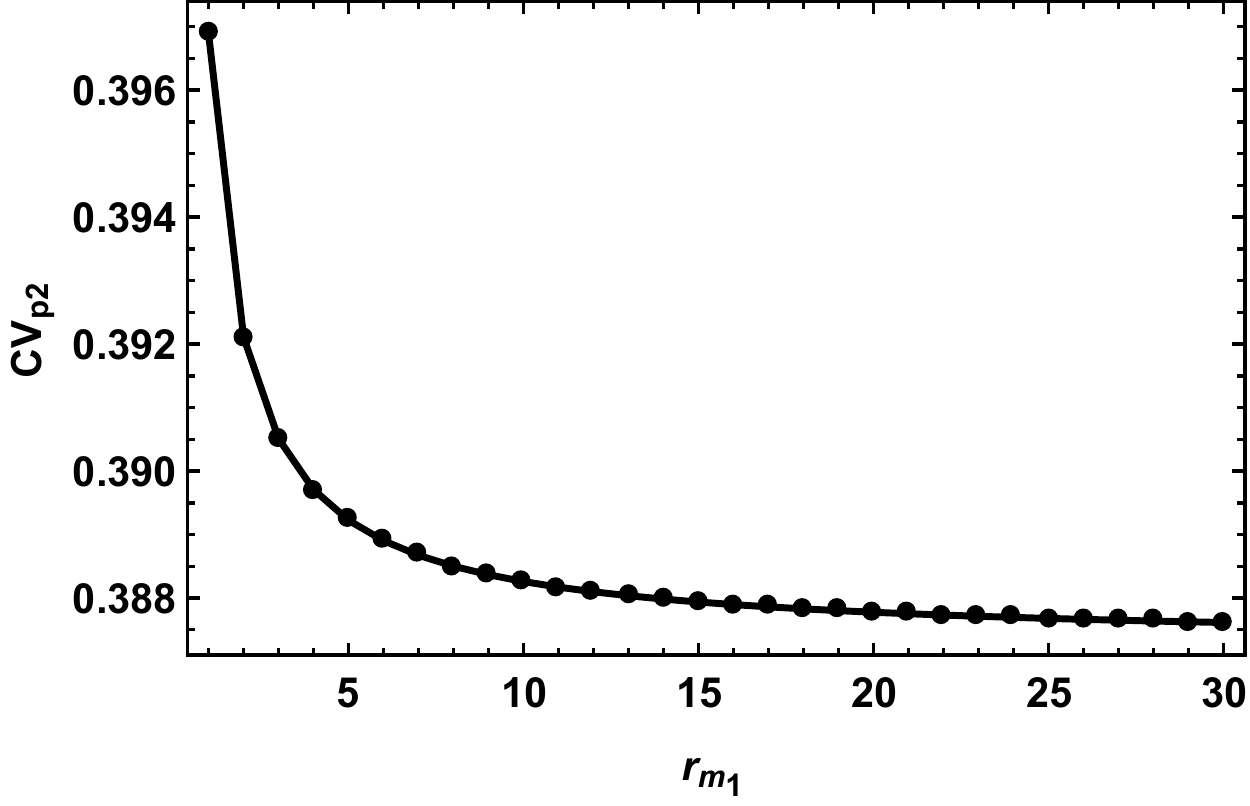}
	\caption{Coefficient of variation for the target protein in the case of sFFL with $\sigma_1=\sigma_2=0$.  $r_{m_1}$ represents 
		$m_1$ synthesis rate. Here $r_s=1$ and the  other parameter values are same as those in figure \ref{fig:cvp2-rm1}. }
	\label{fig:noise-no-ct}
\end{figure}

In order to see the effect of cross-talk, we have plotted the  coefficient of variation for the target protein 
with $r_{m_1}$ in the main text. In the absence of cross-talk (i.e. for $\sigma_1=\sigma_2=0$), 
the coefficient of variation is significantly different showing no indication of optimal noise attenuation (see figure \ref{fig:noise-no-ct}). 

\section{Noise analysis for  sCN}\label{noisescn}
In the case of sCN,  we follow the same master equation approach as done for sFFL. The master equation  for the 
probability of a given state is based on  the following effective differential  equations  
\begin{eqnarray}
&&\dot{s}=r_s-\gamma_s\, s -g_1\, s\, m_1,\\
&&\dot{m_1}=r_{m_1}-\gamma_{m_1}\, m_1-d_1 \, s\, m_1-g_1\, s\, m_1,\\
&&\dot{p_1} =r_{p_1}' s\, m_1-\gamma_{p_1}\, p_1,\\
&&\dot{m_2}=\frac{r_{m_2}\,k_c\, p_1}{1+k_c\, p_1}-\gamma_{m_2}\,  m_2,\\
&&\dot{p_2}=r_{p_2}\, m_2 -\gamma_{p_2}\,  p_2.
\end{eqnarray}

	\begin{figure}[h]
	\centering
	\includegraphics[width=\linewidth]{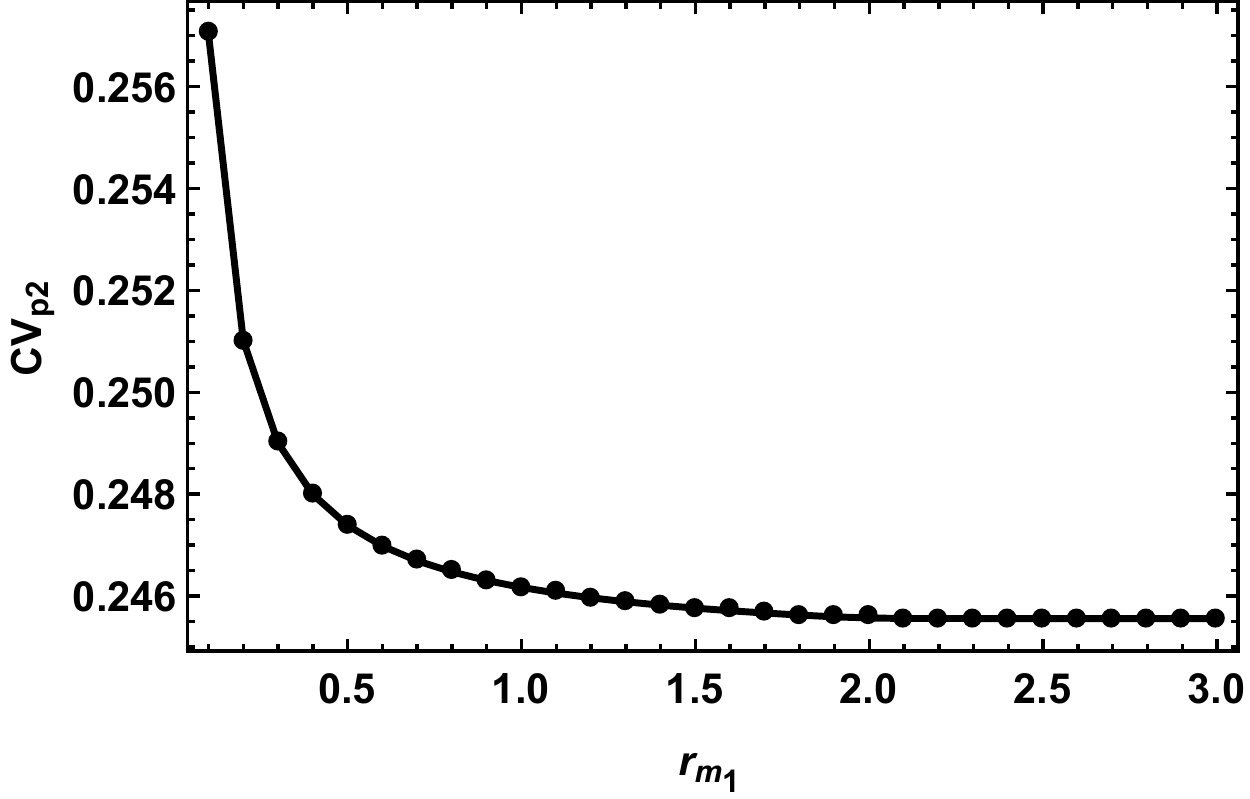}
	\caption{Coefficient of variation for the target protein plotted with $r_{m_1}$ in sCN. Here $r_s=1$ and all other parameter values are same as those in figure \ref{fig:cvp2-rm1}. Further, in the case of sCN, $k_2^+=k_2^-=\sigma_2=\kappa_2=0$. }
	\label{fig:cvscn}
	\end{figure}
In figure  \ref{fig:cvscn}, we plot   the  coefficient of variation $CV_{p_2}=(G_{55}+G_5-G_5^2)^{1/2}/G_5$  with  $r_{m_1}$.   No minimum 
in the coefficient of variation is found in this case. Here $r_s=1$ and all other parameter values are same as those in figure \ref{fig:cvp2-rm1}. 

In the following, we present  first and second moments necessary for obtaining the coefficient of variation for the 
sRNA-driven cascade network. As before, we use Gaussian approximation to express the third moments in terms of various 
first and second  moments. 
	\begin{align}
	&G_1 = \frac{r_s - g_1\,G_{12}}{\gamma_s}\\
	&G_2 = \frac{r_{m_1}  - (g_1+ d_1) G_{12}}{\gamma_{m_1}}\\
	&G_3 = \frac{r_{p_1}' G_{12}}{\gamma_{p_1}}\\ 
	&G_4 = \frac{r_{m_2}^0 + r_{m_2}^1\, G_3 }{\gamma_{m_2}}\\
	&G_5 = \frac{r_{p_2} G_{4}}{\gamma_{p_2}}
	\end{align}
	\begin{align}
	&G_{11} = \frac{r_s\, G_1 - g_1\,G_{112}}{\gamma_s}\\
	&G_{22} = \frac{r_{m_1} G_2 - (g_1+ d_1) G_{122} }{\gamma_{m_1}}\\
	&G_{33} = \frac{r_{p_1}' G_{123}}{\gamma_{p_1}}\\
	&G_{44} = \frac{r_{m_2}^0 G_4 + r_{m_2}^1 G_{34} }{\gamma_{m_2}}\\ 
	&G_{55} = \frac{r_{p_2} G_{45}}{\gamma_{p_2}}\\
	&\scalebox{1.25}
	{$G_{12} = \frac{r_s G_2 + r_{m_1} G_1 - g_1(G_{112} + G_{122}) - d_1G_{112}}{\gamma_s + \gamma_{m_1} + g_1+ d_1}$}\\
	&\scalebox{1.25}
	{$G_{13} = \frac{r_s G_3 + r_{p_1}' ( G_{12} + G_{112}) - g_1\,G_{123}}{\gamma_s + \gamma_{p_1}}$}\\
	&\scalebox{1.1}
	{$G_{14} = \frac{r_s G_4 + r_{m_2}^0 G_1 + r_{m_2}^1  G_{13} - g_1\,G_{124}}{\gamma_s + \gamma_{m_2} }$}\\
	&\scalebox{1.25}
	{$G_{15} = \frac{r_s G_5 + r_{p_2} G_{14} - g_1\,G_{125} }{\gamma_s + \gamma_{p_2}}$}\\
	&\scalebox{1.25}
	{$G_{23} = \frac{r_{m_1} G_3 + r_{p_1}' ( G_{12} + G_{122}) - (g_1+ d_1) G_{123}}{\gamma_{m_1} + \gamma_{p_1}}$}\\
	&\scalebox{1.25}
	{$G_{24} = \frac{r_{m_1} G_4 + r_{m_2}^0 G_2 + r_{m_2}^1 G_{23} - (g_1+  d_1) G_{124} }{\gamma_{m_1} + \gamma_{m_2}}$}\\
	&\scalebox{1.25}
	{$G_{25} = \frac{r_{m_1} G_5 + r_{p_2} G_{24} - (g_1+ d_1) G_{125}}{\gamma_{m_1} + \gamma_{p_2} }$}\\
	&\scalebox{1.25}
	{$G_{34} = \frac{r_{p_1}' G_{124} + r_{m_2}^0  G_3 + r_{m_2}^1 ( G_3 + G_{33})}{\gamma_{p_1} + \gamma_{m_2}}$}\\
	&\scalebox{1.25}
	{$G_{35} = \frac{r_{p_1}' G_{125} + r_{p_2} G_{34}}{\gamma_{p_1} + \gamma_{p_2} }$}\\
	&\scalebox{1.25}
	{$G_{45}=\frac{r_{m_2}^0 G_5 + r_{m_2}^1 G_{35} + r_{p_2}( G_{4} + G_{44})}{\gamma_{m_2} + \gamma_{p_2}}$}
	\end{align}

\begin{widetext}
	\section{Stochastic  simulations}\label{reactions}
	The reactions considered for the stochastic simulations  and the corresponding rates are listed below. 
		\begin{eqnarray}
	&&\schemestart $\phi$   \arrow{<=>[$ r_s $][$\gamma_{s}$]}  $s$ \schemestop \quad ({\rm synthesis\, and \, degradation\, of\,  sRNA;}\,  r_s= variable,  \gamma_s=0.001\ (\rm s^{-1}) )
	\\
	&&\schemestart $\phi$   \arrow{<=>[$ r_{ m_1} $][$\gamma_{{ m_1}}$]}  $ m_1$ \schemestop \quad ({\rm synthesis\, and \, degradation\ of\  mRNA1;}\  r_{ m_1}=variable, \gamma_{ m_1}=0.001\ (\rm s^{-1}) )
	\\
	&&\schemestart $m_1$ + $s$  \arrow{<=>[$ k_1^+ $][$k_1^-$]}  $ c_1$ \schemestop 
	\begin{cases}
	& {\rm association\, and \, dissociation\ of\ sRNA\mbox{-}mRNA1\ complex};\\
	&  k_1^+=0.1\ {\rm (molecules^{-1}.\ s^{-1})},\ k_1^-=0.05\ (\rm s^{-1}).
	\end{cases}
	\\
	&&\schemestart $c_1$ \arrow{->[$\sigma_{1}$]}  $\phi$\ \schemestop\quad ({\rm stoichiometric\ degradation\ of\ sRNA\mbox{-}mRNA1\ complex \ (c_1);}\ \sigma_{1}=0.005\ (\rm s^{-1})) 
	\\ 
	&&\schemestart $c_1$ \arrow{->[$\kappa_{1}$]}  $s$\ \schemestop\quad ({\rm catalytic\ degradation\ of\ sRNA\mbox{-}mRNA1\ complex\  (c_1);}\ \kappa_{1}=0.005\ (\rm s^{-1})) 
	\\
	&&\schemestart $c_1$ \arrow{->[$r_{p_1}$]} $p_1$ + $c_1$ \schemestop\quad ({\rm translation\ and\  synthesis \ of\   protein,\ p_1};\  r_{p_1}=0.01\  ({\rm molecules.\ s^{-1})})  
	\\
	&&\schemestart $p_1$ \arrow{->[$\gamma_{p_1}$]}  $\phi$\ \schemestop\quad ({\rm degradation\ of\ protein,\  p_1;}\ \gamma_{p_1}=0.001\ (\rm s^{-1}))   
	\\
	&&\schemestart $p_1$ + $G_{p_2}$ \arrow{<=>[$k_c^+$][$k_c^-$]}   $G_{p_2}^*$\ \schemestop
	\begin{cases}
	&{\rm transcriptional\, activation\, of\,  gene\, synthesising\ protein, p_2;}\,\\
	& k_c^+=0.2\,  {(\rm molecules^{-1}.\, s^{-1})}, \ k_c^-=2\,  {(\rm s^{-1})}
	\end{cases}
	\\
	&&\schemestart $G_{p_2}^*$ \arrow{->[$r_{m_2}$]} $m_2$ +  $G_{p_2}^*$\schemestop\quad ({\rm synthesis\ of\  mRNA2; }\  r_{m_2}= 0.01\  (\rm molecules.\ s^{-1}))  
	\\
	&&\schemestart $m_2$ + $s$  \arrow{<=>[$ k_2^+ $][$k_2^-$]}  $ c_2$ \schemestop 
	\begin{cases}
	& {\rm association\, and \, dissociation\ of\ sRNA\mbox{-}mRNA2\ complex};\\
	&  k_2^+=0.1\ {\rm (molecules^{-1}.\ s^{-1})},\ k_2^-=0.05\ (\rm s^{-1}).
	\end{cases}
	\\
	&&\schemestart $c_2$ \arrow{->[$\sigma_{2}$]}  $\phi$\ \schemestop\quad ({\rm stoichiometric\ degradation\ of\ sRNA\mbox{-}mRNA2 \  complex;}\ \sigma_{2}=0.005\ (\rm s^{-1})) 
	\\ 
	&&\schemestart $c_2$ \arrow{->[$\kappa_{2}$]}  $s$\ \schemestop\quad ({\rm catalytic\ degradation\ of\ sRNA\mbox{-}mRNA2\  complex;}\ \kappa_{2}=0.005\ (\rm s^{-1})) 
	\\
	&&\schemestart $m_2$ \arrow{->[$\gamma_{m_2}$]} $\phi$\schemestop\quad  ({\rm degradation\ of\ mRNA2;}\ 
	\gamma_{m_2}=0.001\ (\rm s^{-1})) 
	\\
	&&\schemestart $c_2$ \arrow{->[$r_{p_2}$]} $p_2$ + $c_2$ \schemestop\quad ({\rm translation\ and\  synthesis \ of\   protein,\ p_2};\  r_{p_2}=0.01 \  ({\rm molecules.\ s^{-1})})  
	\\
	&&\schemestart $p_2$ \arrow{->[$\gamma_{p_2}$]}  $\phi$\schemestop\quad ({\rm degradation\ of\ protein,\ p_2}; \ \gamma_{p_2}=0.001\ (\rm s^{-1}))
	\end{eqnarray}
\end{widetext}

\end{document}